\documentclass[showpacs,aps,preprintnumbers,prb,amsmath,amssymb,twocolumn,floatfix]{revtex4-1}
\usepackage{graphicx}
\usepackage{color}
\usepackage[normalem]{ulem}
\usepackage{amssymb}
\usepackage{amsmath}
\usepackage[normalem]{ulem}
\usepackage{hyperref}
\DeclareGraphicsExtensions{.pdf,.eps,.png,.jpg,.mps}

\begin{document}

\title{Effect of atomic disorder and Ce doping on superconductivity of Ca$_{3}$Rh$_4$Sn$_{13}$:\\ Electric transport properties under  high  pressure. }
\author{A.~\'{S}lebarski$^{1,2}$, J. Goraus$^{1}$, M. M. Ma\'{s}ka$^{1}$, P. Witas$^{1}$, M. Fija\l kowski$^{1}$, C. T. Wolowiec$^{3}$, Y. Fang$^{3,4}$, and M. B. Maple$^{3}$}
\affiliation{
$^{1}$Institute of Physics,
University of Silesia, Uniwersytecka 4, 40-007 Katowice, Poland\\
$^{2}$Centre for Advanced Materials and Smart Structures, 
Polish Academy of Sciences, Ok\'{o}lna 2, 50-950 Wroc\l aw, Poland\\
$^{3}$Department of Physics, University of California, San Diego, La Jolla,
California 92093, USA\\
$^{4}$Materials Science and Engineering Program, University of California, San Diego, La Jolla, California 92093, USA}

\begin{abstract}

We report the observation of a superconducting state below $\sim 8$ K coexistent with a spin-glass state caused by atomic disorder 
in Ce substituted Ca$_3$Rh$_4$Sn$_{13}$. Measurements of specific heat, resistivity, and magnetism reveal the existence of inhomogeneous superconductivity in samples doped with Ce with superconducting critical temperatures $T_c$ higher than those observed in the parent compound. For Ca$_3$Rh$_4$Sn$_{13}$, the negative value of the change in resistivity $\rho$ with pressure $P$, $\frac{d\rho}{dP}$ correlates well with the calculated decrease in the density of states (DOS) at the Fermi energy with $P$. Based on band structure calculations performed under pressure, we demonstrate how the change in DOS  would affect $T_c$ of Ca$_3$Rh$_4$Sn$_{13}$ under negative lattice pressure in samples that are strongly defected by quenching.

\end{abstract}


\maketitle

\section{Introduction}

The interplay of magnetic order and superconductivity  is a key topic of current research of strongly correlated  heavy fermion compounds, high-$T_c$ cuprates as well as iron-based pnictide or chalcogenide superconductors. 
The destruction of superconductivity at the onset of magnetic order has been well known; however,  there are an increasing number of new superconductors which are found experimentally to be closely linked to magnetism. Analysis of the magnetic ordering and its coexistence with superconductivity may help to explain the pairing mechanism in high-$T_c$ and unconventional superconductors. 
In particular, very recent reports demonstrate coexistence of superconducting and spin-glass-like states in EuFe$_2$(As$_{1-x}$P$_x$)$_2$ \cite{Zapf2013}, and Ca$_{0.9}$Ce$_{0.1}$Fe$_2$As$_2$ \cite{Nadeem2015}, which suggests an important  role of atomic disorder. 

There are a number of strongly correlated superconductors including Bi$_2$Sr$_2$CaCu$_2$O$_{8+x}$, PrOs$_4$Sb$_{12}$ \cite{Maple2002,Vollmer2003}, CePt$_3$Si \cite{Kim05,Takeuchi07}, and CeIrIn$_5$ \cite{Bianchi01}, in which the nanoscale electronic disorder is responsible for an increase in $T_c$. Similar behavior has been found in La$_3$Rh$_4$Sn$_{13}$ \cite{Slebarski2014a} and La$_3$Ru$_4$Sn$_{13}$ \cite{Slebarski2015}, where nanoscale electronic disorder leads to the increase of $T_c$. 
Regarding the filamentary superconductivity observed in known granular superconductors, it has been reported \cite{Xiao2012}, for example, that the filamentary superconductivity in CaFe$_2$As$_2$ suggests that even nominally pure stoichiometric 
compounds can spontaneously become electronically inhomogeneous at the nanoscale. In the similar compound, 
BaFe$_2$As$_2$, \cite{Dioguardi} a spin-glass-like antiferromagnetic state was found to be coexist  with superconductivity along domain walls.
It is widely believed that disorder plays a key role in the new superconductors. In particular, such disorder can have noticeable effects on  $T_c$.
In this study, we report an increase in $T_c$ when Ce is substituted for Ca in the skutterudite-related Ca$_3$Rh$_4$Sn$_{13}$ superconductor. We argue that it can be a result of the disorder induced by the substitution.
The results will contribute to a better understanding of the complex low temperature properties observed in novel superconducting strongly correlated electron systems.

\section{Experimental details}   

\begin{figure}[h!]
\includegraphics[width=0.48\textwidth]{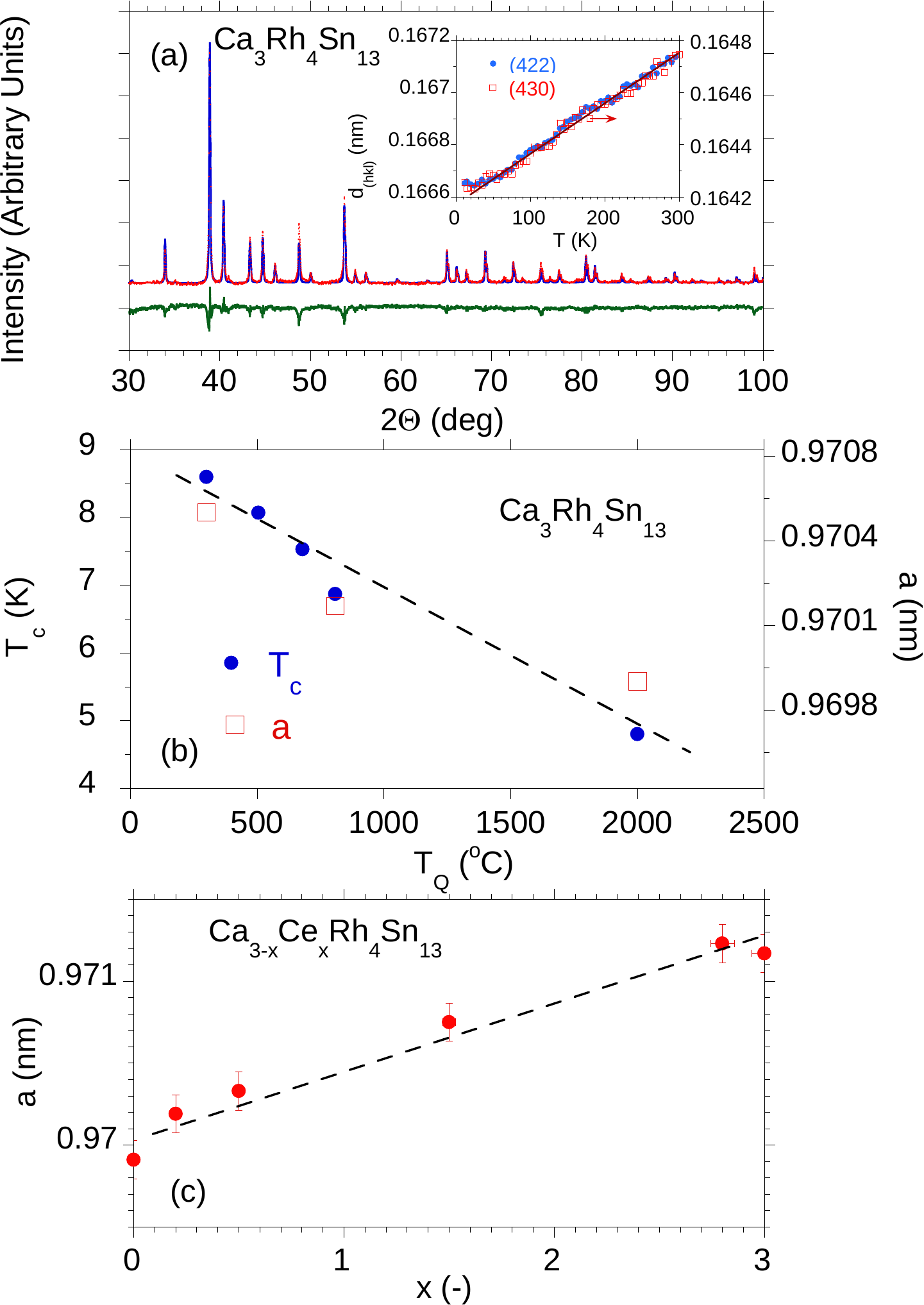}
\caption{\label{fig:Fig1}
($a$) X-ray diffraction pattern for the Ca$_{3}$Rh$_{4}$Sn$_{13}$ compound. The green line represents the differences between the blue and red line peak shapes. Inset: The temperature change of the (422) and (430) XRD lines between 12 K and 300 K. There was no change of $d_{hkl}$ vs $T$, characteristic of the structural phase transition. ($b$) $T_c$ and the lattice parameter $a$ for Ca$_{3}$Rh$_{4}$Sn$_{13}$ plotted against different quenching temperatures $T_Q$. The data at $T_Q=505,\: 679$, and 804$^o$C are taken from Ref. \onlinecite{Westerveld1987}. Temperature $T_c=4.8$ K and the room temperature lattice parameter $a=9.6991$ \AA\ for the arc-melted sample are expected at $T_Q$ $\sim$ 2000 K. ($c$) The lattice parameter, $a$, plotted against Ce concentration, $x$, for the  Ca$_{3-x}$Ce$_x$Rh$_4$Sn$_{13}$ system. The dotted line is a linear fit to the data. 
}
\end{figure}

\begin{figure}[h!]
\includegraphics[width=0.48\textwidth]{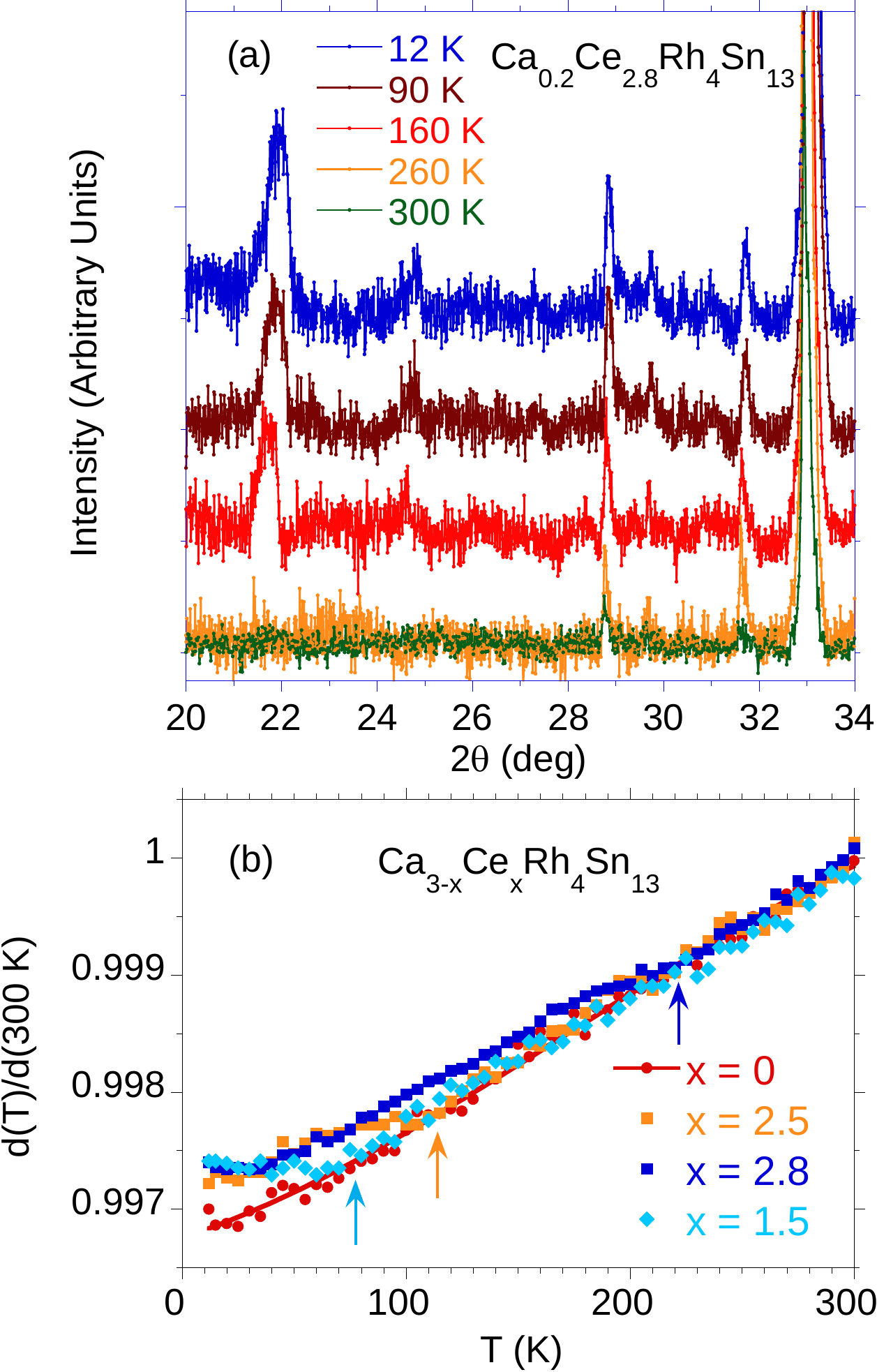}
\caption{\label{fig:XRD}
($a$) X-ray diffraction of Ca$_{0.2}$Ce$_{2.8}$Rh$_4$Sn$_{13}$ with Cu $K_{\alpha}$ radiation at different temperatures for small $2\theta$s to exhibit additional superlattice reflections at $T=160$ K and below. ($b$) The temperature change of the distance $d(T)$ in different compounds Ca$_{1-x}$Ce$_{x}$Rh$_4$Sn$_{13}$, normalized to the $d$ value at $T=300$ K between (422) crystallographic planes, obtained from Bragg equation $\lambda=2d \sin \theta(T)$ for diffraction line (422). Arrows indicate the temperature of the structural distortion ($T_D$) for the components $x=1.5$, 2.5, and 2.8. There is a change of intensity of the XRD diffraction lines at $T_D$, not shown here (c.f. Ref. \onlinecite{Slebarski2013a}). In panel ($a$), for $T\leq 160$ K the intensity of the superlattice reflection at $2\theta \approx 22$ deg shows linear increase vs decreasing $T$.
}
\end{figure}

\begin{figure}[h!]
\includegraphics[width=0.48\textwidth]{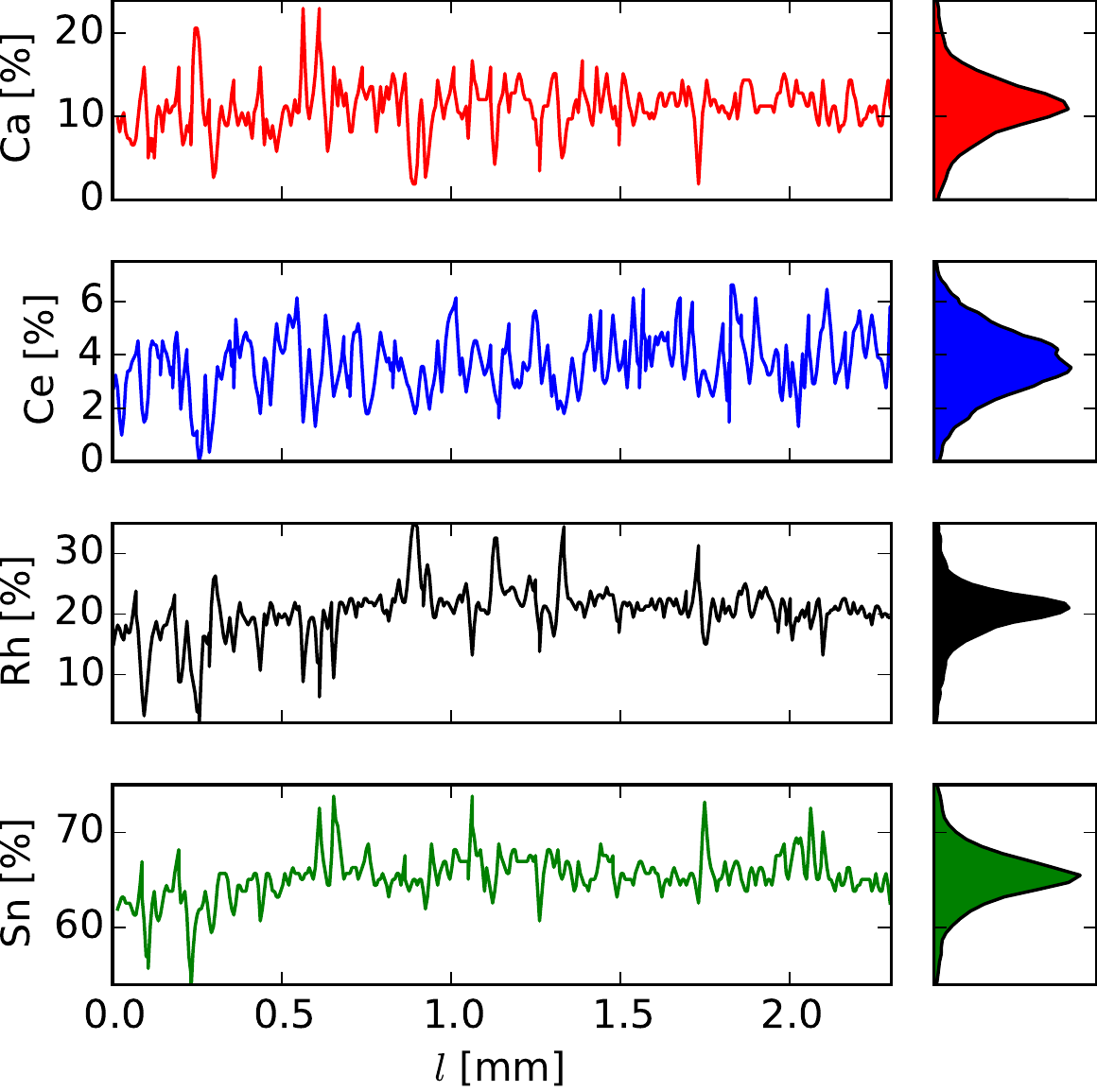}
\caption{\label{fig:Variation}
  Variations in stoichiometry of Ca$_{2.4}$Ce$_{0.6}$Rh$_{4}$Sn$_{13}$ over the length of the sample. The right panels
  show corresponding histograms.}
\end{figure}

The Ca$_{3}$Rh$_{4}$Sn$_{13}$ and Ce$_{3}$Rh$_{4}$Sn$_{13}$ polycrystalline samples were prepared by arc melting the constituent elements on a water cooled copper hearth in a high-purity argon atmosphere with an Al getter. 
The Ca$_{3-x}$Ce$_x$Rh$_4$Sn$_{13}$ alloys were then prepared by
diluting the parent compounds with nominal compositions of Ce and Ca which were then annealed  
at 870$^{o}$C for 2 weeks. All samples were examined by x-ray diffraction (XRD) analysis and found to have a cubic structure (space group Pm$\bar{3}$n) \cite{Remeika80}. 
XRD data were refined using FULPROF program \cite{fulprof} with pseudo-Voight line shape, yielding the profile $R$-factor $R_{p}<4$, the weighted profile $R$-factor $R_{wp}<5$, and $R_{exp}\approx 0.5$. Figure \ref{fig:Fig1}$a$ shows a representative XRD pattern for 
Ca$_{3}$Rh$_{4}$Sn$_{13}$ together with its Rietveld refined pattern.  From the XRD patterns measured at $T=300$ K, 160 K and 12 K we identified the cubic structure of  Yb$_{3}$Rh$_{4}$Sn$_{13}$-type
for Ca$_{3}$Rh$_{4}$Sn$_{13}$. 
No evidence of simple cubic-to-bcc structural distortion was found in Ca$_{3}$Rh$_{4}$Sn$_{13}$, while the low-temperature XRD data show superlattice reflections for  $x \geq 1.5$ components of Ca$_{3-x}$Ce$_x$Rh$_{4}$Sn$_{13}$  in the low $2\theta$ region (see Fig. \ref{fig:XRD}$a$), interpreted (c.f. Refs. \cite{Klintberg2012,Goh2015}) as a structural transition from simple cubic to I$\bar{4}$3d bcc structure, involving crystallographic cell doubling. Fig. \ref{fig:XRD}$b$ exhibits the abnormal volume change at the temperature of structural distortion found for $x\geq 1.5$, which is, however,  not observed for the parent compound.
The refined lattice parameters shown in Fig. \ref{fig:Fig1}$c$  and corresponding atomic positions were used in our band structure calculations.  
The lattice parameters of the Ca$_{3-x}$Ce$_x$Rh$_4$Sn$_{13}$ alloys measured at room temperature follow Vegard's law, which suggests good sample quality and stoichiometry of the components $x$. 
Stoichiometry and homogeneity were checked by the electron microprobe technique (scanning microscope JSM-5410) and by XPS analysis. Deviations from the nominal composition were small. The experimentally obtained compositions from the total surface of the sample were very close to the assumed one; e.g., for Ca$_{2.6}$Ce$_{0.6}$Rh$_{4}$Sn$_{13}$ with the assumed atomic concentration ratio 12:3:20:65, respectively  we determined 12.05 at.\% Ca, 3.04 at.\% Ce, 19.60 at.\% Rh, and 65.31 at.\% Sn. The variations in stoichiometry over the length of the sample are shown for this sample in Fig.~\ref{fig:Variation}. 
Local fluctuations in composition were observed in nano-scale for all components of the alloy, however, the greatest one exist for Ce, which may explain the strong disorder induced by doping. We observed similar fluctuations in the composition for the parent compound, which signals similar site disorder induced by rapid quenching from high temperature in the system of Ca$_{3-x}$Ce$_x$Rh$_{4}$Sn$_{13}$ alloys.

The difference spectrum shown in Fig. \ref{fig:Fig1}$a$ (green line) results from the differences in peak shape, which we attribute to atomic disorder in Ca$_{3}$Rh$_{4}$Sn$_{13}$. It was reported \cite{Westerveld1987,Westerveld1989}, that  Ca-Sn1 anti-site defects generated at high temperatures and then frozen-in by rapid quenching to room temperature are responsible either for the degradation of the superconducting transition temperature $T_c$ or a reduction in the lattice parameter $a$ in Ca$_{3}$Rh$_{4}$Sn$_{13}$. Figure \ref{fig:Fig1}$b$ exhibits the $T_c$ vs $T_Q$ (the temperature from which the samples were quenched) data taken from Ref. \onlinecite{Westerveld1987}.
In the quenching temperature range $\sim 810-300$ $^{o}$C, $T_c$ of Ca$_{3}$Rh$_{4}$Sn$_{13}$
is lowered from 8.4 K in the as-grown sample down to 6.9 K in the sample which was quenched at $\sim 810$ K \cite{Westerveld1987}. Hence, the value of $T_c\sim 4.8$ K and the lattice parameter $a=9.6991$ \AA\ for the sample melted in arc (shown in the figure)
can be explained by rapid quenching at $T_Q$ of about $2000-3000$ K.\\

\indent Electrical resistivity $\rho$ at ambient pressure
was investigated by a conventional four-point ac technique using a Quantum Design Physical Properties Measurement System (PPMS). 
Measurements of $\rho$ under pressure were performed in a beryllium-copper, piston-cylinder clamped cell (for details, see Ref. \onlinecite{Slebarski2012}).\\ 
\indent Specific heat $C$ was measured in the temperature range $0.4-300$ K and in external magnetic fields up to 9 T using a Quantum Design PPMS platform. The dc magnetization $M$ and (dc and ac) magnetic susceptibility $\chi$ were obtained using a commercial superconducting quantum interference device magnetometer from 1.8 K to 300 K in magnetic fields up to 7 T.\\
\indent The XPS spectra were obtained
with monochromatized Al $K_{\alpha}$ radiation at room temperature
using a PHI 5700 ESCA spectrometer. The polycrystalline sample was broken under high vacuum better than $6\times 10^{-10}$ Torr immediately before taking a spectrum.\\
\indent The band structure calculations were accomplished using fully relativistic full potential local orbital method
(FPLO9-00-34 computer code \cite{fplo}) within the local spin density approximation (LSDA) as well as ELK FP-LAPW/APW+lo code \cite{elk}. The exchange correlation potential V$_{xc}$ was used in the form proposed by Perdew-Wang \cite{pw92} in both cases.
The number of k-points in the irreducible wedge of Brillouin Zone was 80.
The results obtained from both methods were accomplished for the same V$_{xc}$, and as expected were essentially the same.
The ELK-code was used for accurate calculations of the electron localization function (ELF), whereas the FPLO method was used to study the pressure influence as it is much faster due to a more advanced mixing scheme.

\section{Results and discussion}

\subsection{Atomic disorder and superconductivity in Ca$_{3}$Rh$_{4}$Sn$_{13}$ revisited}
\subsubsection{Magnetic properties of Ca$_{3}$Rh$_{4}$Sn$_{13}$ doped with Ce, evidence of short range magnetic order and the superconducting state}

\begin{figure}[h!]
\includegraphics[width=0.48\textwidth]{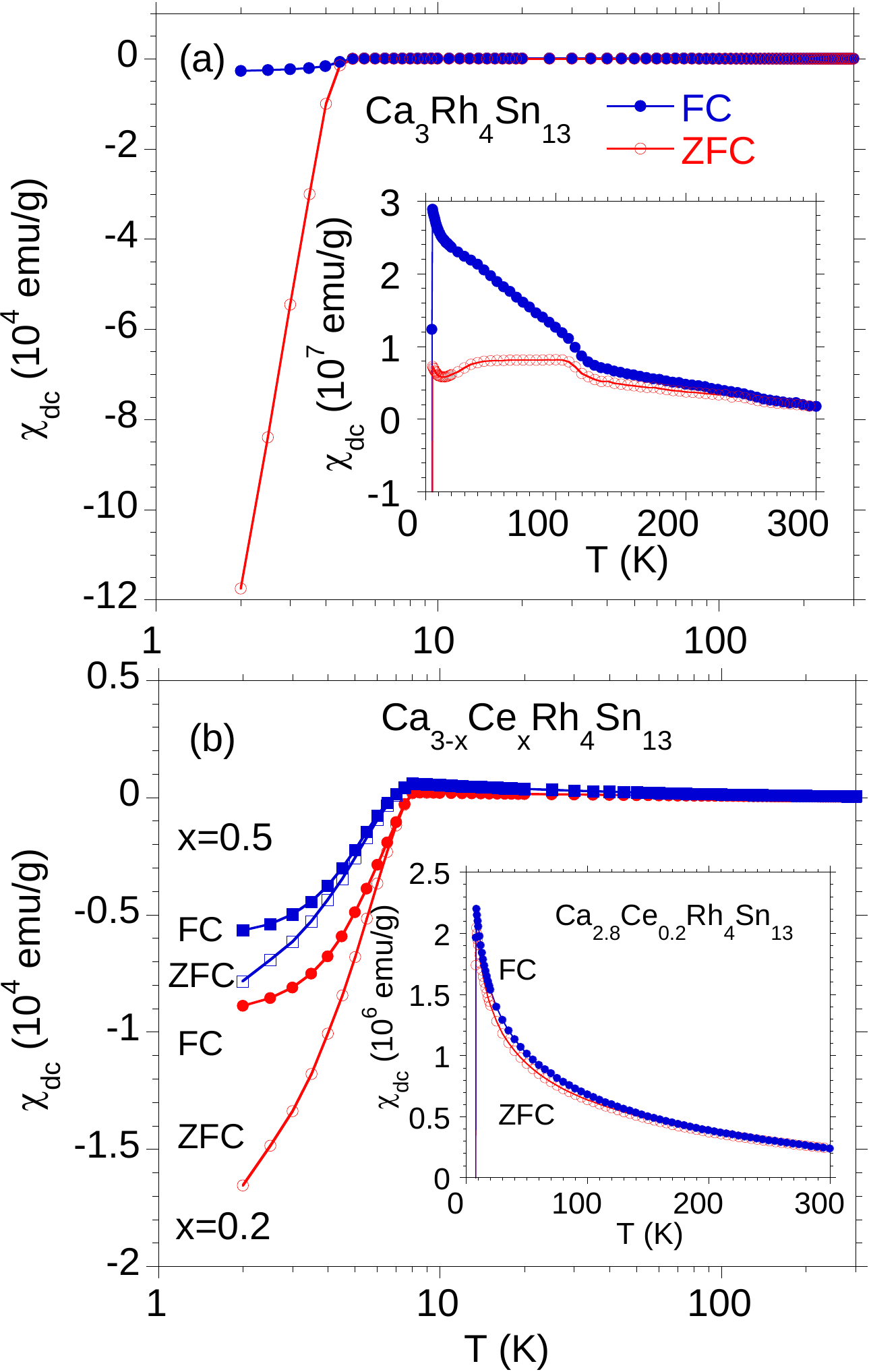}
\caption{\label{fig:Fig2}
Magnetic susceptibility (dc) for Ca$_{3-x}$Ce$_x$Rh$_{4}$Sn$_{13}$ in a field-cooled (FC) and zero-field-cooled (ZFC) experiment with an applied field 0.05 T. The data are shown for Ca$_{3}$Rh$_{4}$Sn$_{13}$ ($a$) and for the compounds with $x=0.2$ and $0.5$ ($b$). The insets exhibit details in the extended $\chi$ axes.
}
\end{figure}

\begin{figure}[h!]
\includegraphics[width=0.47\textwidth]{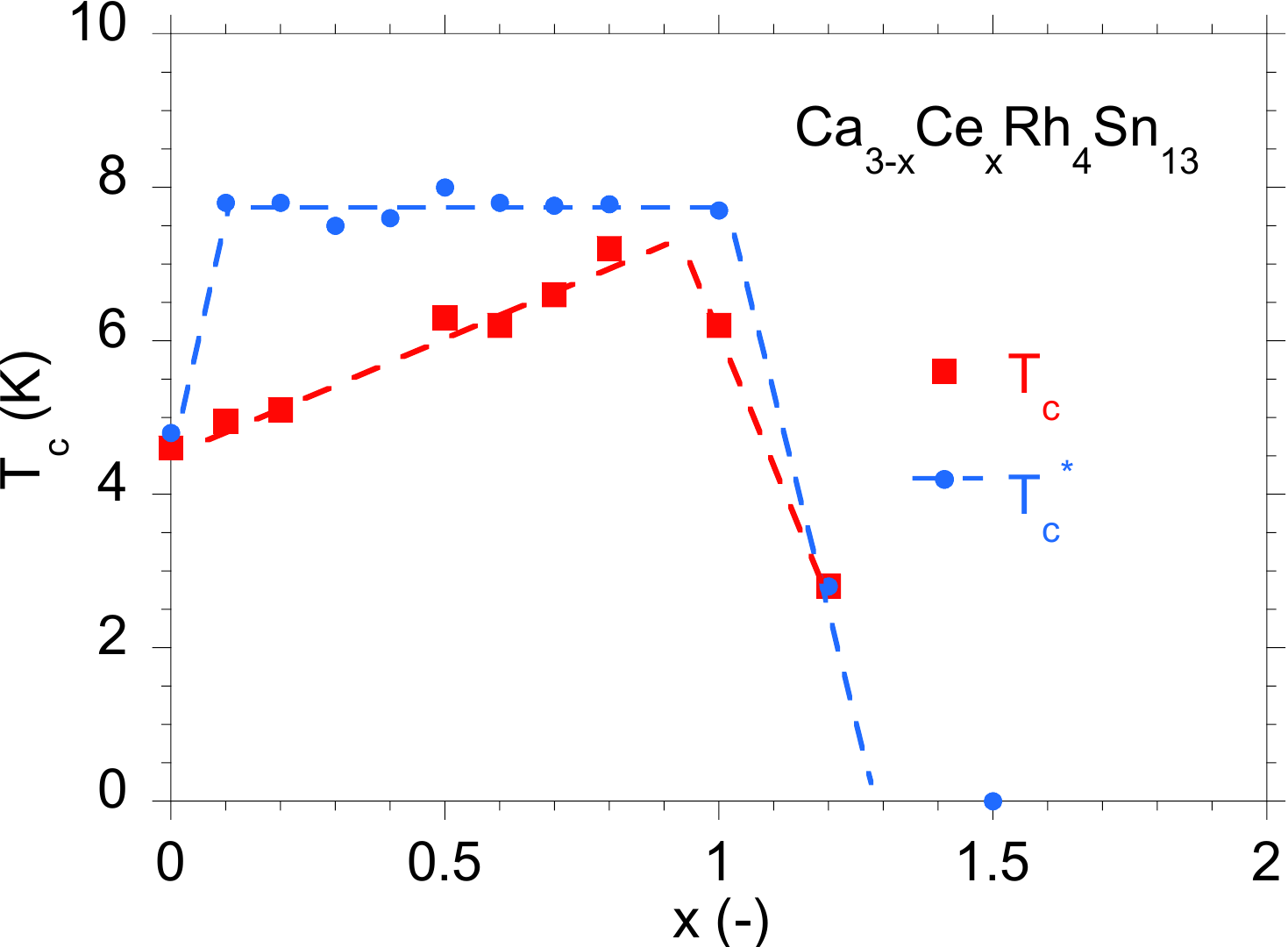}
\caption{\label{fig:dome}
  Superconducting critical temperature $T_c$ (red points) and $T_c^*$ (blue points) of Ca$_{3-x}$Ce$_x$Rh$_{4}$Sn$_{13}$ for $x$ from 0 to 1.2 obtained from $\chi_{ac}$.
  An increase in Ce doping leads to an increase of the lattice constant (see Fig. \ref{fig:Fig1}c), suggesting an effective {\it chemical pressure} for the $x \leq 0$, undoped system (discussed later in the text).}
\end{figure}

\begin{figure}[h!]
\includegraphics[width=0.48\textwidth]{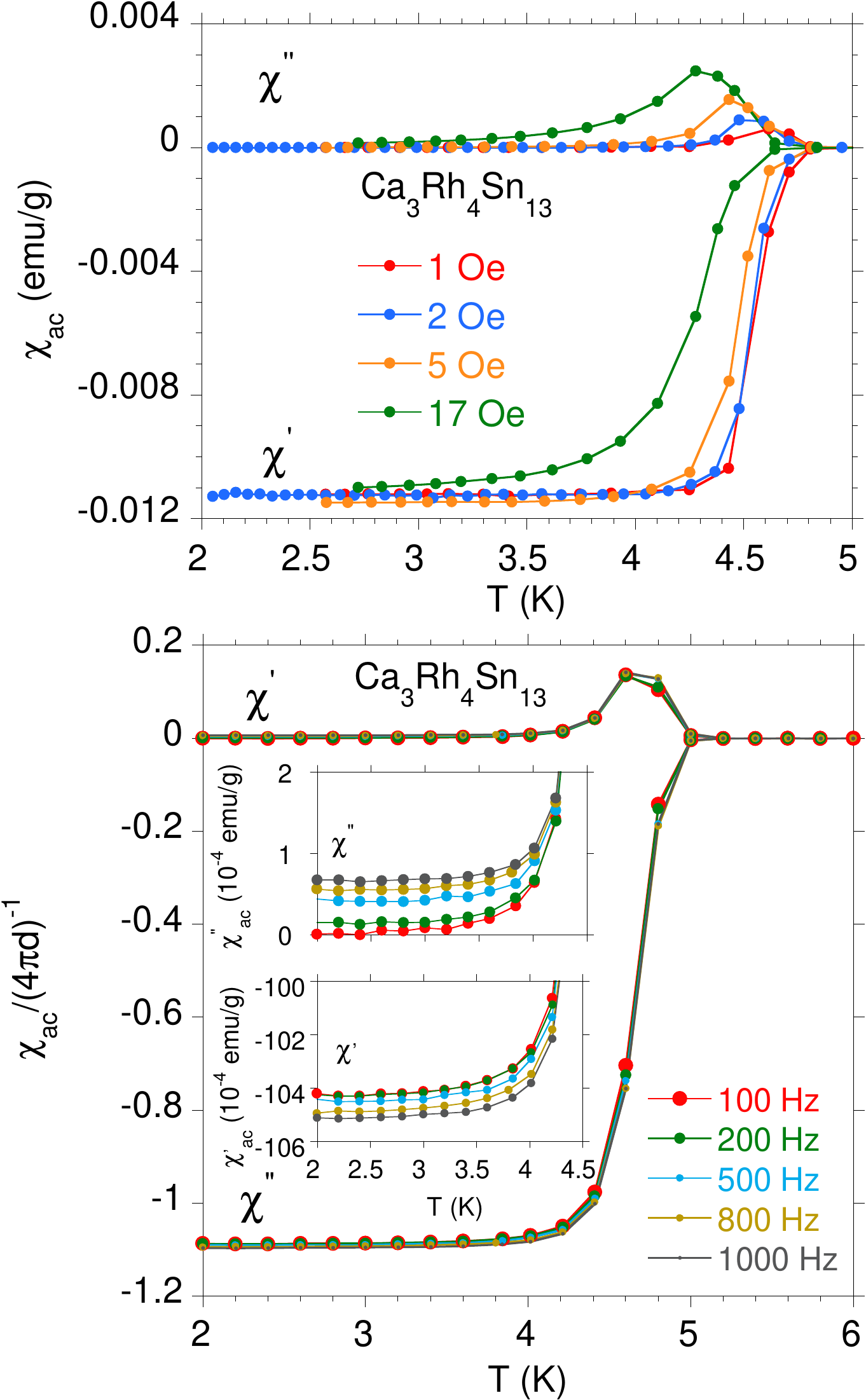}
\caption{\label{fig:Fig3}
The real and imaginary components of the ac magnetic susceptibility, $\chi^{'}$ and $\chi^{''}$, for Ca$_{3}$Rh$_{4}$Sn$_{13}$, as a function of temperature measured in various magnetic fields ($a$) and at different frequencies ($b$) in a field $B=4$ Oe. 
}
\end{figure}

\begin{figure}[h!]
\includegraphics[width=0.48\textwidth]{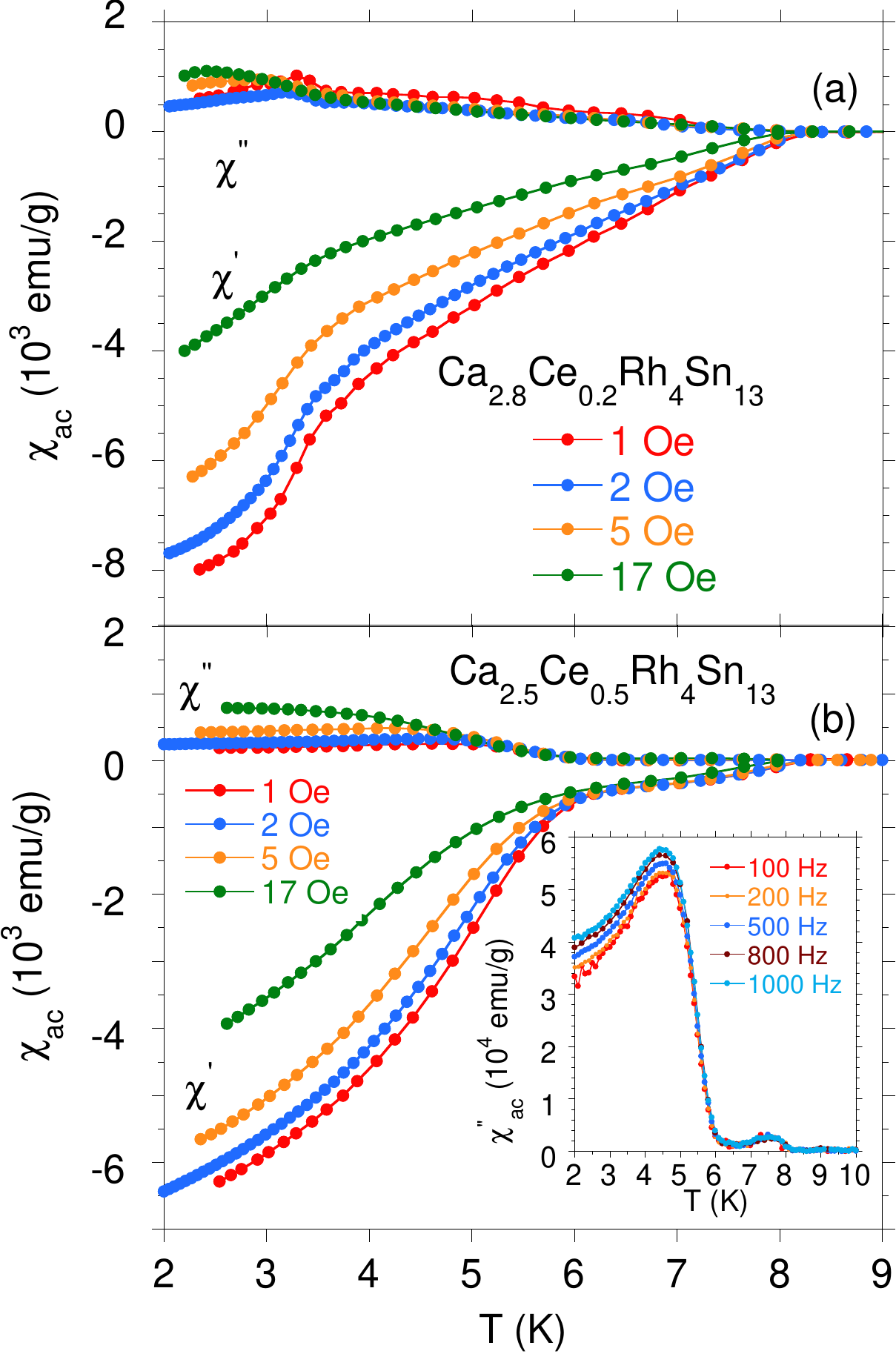}
\caption{\label{fig:Fig4}
The real and imaginary components of the ac magnetic susceptibility, $\chi^{'}$ and $\chi^{''}$, for Ca$_{2.8}$Ce$_{0.2}$Rh$_{4}$Sn$_{13}$ ($a$) and Ca$_{2.5}$Ce$_{0.5}$Rh$_{4}$Sn$_{13}$ ($b$), as a function of temperature measured in various magnetic fields. The inset to panel ($b$) shows how $\chi^{''}$ depends on frequency for Ca$_{2.5}$Ce$_{0.5}$Rh$_{4}$Sn$_{13}$ with evidence of the spin-glass state.
}
\end{figure}

Ca$_{3}$Rh$_{4}$Sn$_{13}$ has been reported to be a paramagnetic \cite{Tomy1997} superconducting compound with a $T_c\sim 8.4$ K. However, the superconducting state of this compound was proposed \cite{Westerveld1987} to be strongly dependent on the atomic disorder, which, upon quenching, leads to a significant decrease in $T_c$. On the other hand, there are some examples of analogous skutterudite-related superconductors (La$_{3}$Rh$_{4}$Sn$_{13}$ \cite{Slebarski2014}, La$_{3}$Ru$_{4}$Sn$_{13}$ \cite{Slebarski2015}) which show evidence of nanoscale disorder over length scales similar to the coherence length as a bulk property, leading to an inhomogeneous superconducting state with an enhanced critical temperature $T^{\star}_c>T_c$. 
Here, $T^{\star}_c$ represents a drop of resistivity connected to formation of percolation paths, whereas 
$T_c$, determined from the specific heat, indicates the onset of bulk superconductivity.
This behavior is also observed in other strongly correlated $f$-electron superconductors which
we believe will attract future attention. With this motivation, we present the magnetic investigations of  Ca$_{3}$Rh$_{4}$Sn$_{13}$ substituted  with Ce.
The $\chi$ vs $T$ dc magnetic susceptibility data obtained at 500 Oe in zero field (ZFC) and field cooling (FC) modes shown in
Fig. \ref{fig:Fig2} reveals the onset of diamagnetism and thermal hysteresis associated with the superconducting state below $T_c\approx 4.8$ K for Ca$_{3}$Rh$_{4}$Sn$_{13}$. 
At the concentration of Ce $x=0.1$ $T_c$ jumps to about 8 K and remains at this value for $x$ up to 1. Then, with further
increase of $x$ it drops down and at $x=1.5$ there is no superconductivity. For a wide range of Ce concentrations two superconducting
transitions are observed, what will be discussed later. Fig. \ref{fig:dome} shows how these temperatures depend on $x$. 
Since the critical temperature first increases with increasing x, then reaches maximum at {\it optimal dopping} and then decreases,
it forms a {\it dome} similar to that which is typical for the $high-T_c$ cuprates. Of course, the main difference is that for Ca$_{3}$Rh$_{4}$Sn$_{13}$ doped with Ce, there is no lower critical concentration of Ce and as such the undoped parent compound Ca$_{3}$Rh$_{4}$Sn$_{13}$ is superconducting.
Within the  {\it superconducting dome} there are two effects of disorder on the superconductivity, namely the site disorder induced by rapid quenching from high temperature which decreases $T_c$ in the parent compound and the disorder that results from doping with cerium, which increases superconducting transition. 
 The effect of these two types of disorder is clearly seen in Fig. \ref{fig:dome};  these two types of disorder roughly can  be  distinguished  by microanalysis (c.f. Fig. \ref{fig:Variation} and comment).
Measurements of the dc magnetic susceptibility performed on the parent Ca$_{3}$Rh$_{4}$Sn$_{13}$ sample (cf. inset to Fig. \ref{fig:Fig2}$a$) reveal very weak thermal hysteresis in the ZFC and FC data between $\sim 110$ K and $T_c$. 
This is a significant phenomenon which may be interpreted in several ways. 
One interpretation could be the formation of a charge density wave (CDW) ordering or a structural distortion. There are a few 
examples of compounds that are isostructural to Ca$_{3}$Rh$_{4}$Sn$_{13}$, such as La-based \cite{Liu2013} superconductors or Sr$_3$Ir$_4$Sn$_{13}$ superconductors \cite{Klintberg2012,Fang2014} which exhibit CDW ordering and structural distortions at about $100-140$ K. 
Since the structural distortion has not been observed in Ca$_{3}$Rh$_{4}$Sn$_{13}$ \cite{Goh2015},
a weak spin fluctuation in Ca$_{3}$Rh$_{4}$Sn$_{13}$ could also explain the observed  thermal hysteresis for $T_c<T<\sim 110$ K.  
The high temperature thermal hysteresis observed in $\chi_{dc}$ (Fig. \ref{fig:Fig2}$a$), $\chi_{ac}$ vs. frequency at $T<T_c$ and resistivity measurements (shown later) 
are also suggestive of a {\it granular  effect} \cite{Moseley2015}.
This scenario seems to be very probable and will be discussed. 
In panel ($b$) of Fig. \ref{fig:Fig2}, $\chi$ can be well approximated by the modified Curie-Weiss expression $\chi = \chi_0 + C/(T+\theta_{CW})$ with Curie constant $C=0.807$ K emu/mol$_{Ce}$ for Ce$^{3+}$ ion, $\chi_0 =3.7\times 10^{-4}$ emu/mol$_{Ce}$, and $\theta_{CW}=31$ K, suggesting ferromagnetic correlations of Ce magnetic moments. For the similar isovalent compound Ca$_{3}$Ir$_{4}$Sn$_{13}$, there are a number of 
similar anomalies in the magnetic and electric transport properties at $\sim 38$ K which is well above the superconducting transition; these features were attributed 
to ferromagnetic spin fluctuations coupled to superconductivity \cite{Yang2010,Wang2012}, or a CDW effect \cite{Gerber2013}.
In order to study the complex interplay of magnetism and superconductivity in Ca$_{3-x}$Ce$_x$Rh$_{4}$Sn$_{13}$ with $x\leq 0.5$, we have investigated  the ac mass magnetic susceptibility vs frequency and different ac amplitude of the magnetic field (Figs \ref{fig:Fig3} and \ref{fig:Fig4}). The perfect diamagnetism of the full Meissner state $\chi^{'}=-1/(4\pi d)=9.587\times 10^{-3}$ emu/g for mass density $d=8.3$ g/cm$^3$~ \cite{Haas1999} is reached for Ca$_{3}$Rh$_{4}$Sn$_{13}$ below  $T_c=4.6$ K. The Ce doping and increasing amplitude of the magnetic field systematically reduces the Meissner effect about 73 \% in Ca$_{3-x}$Ce$_x$Rh$_{4}$Sn$_{13}$, which is for the ac magnetic field amplitude of 1 Oe at 2 K for the sample $x=0.5$. 
The insets to Fig. \ref{fig:Fig3} (lower panel) display weak frequency dependence of the real ($\chi^{'}$) and imaginary ($\chi^{''}$) parts of 
ac mass magnetic susceptibility $\chi_{ac}$, which could suggest an atomic disorder effect in Ca$_{3}$Rh$_{4}$Sn$_{13}$, while frequency and field dependencies in $\chi^{'}$ and $\chi^{''}$
depicted in Fig. \ref{fig:Fig4} become apparent 
for spin-glass-like behavior in Ce doped alloys.

\begin{figure}[h!]
\includegraphics[width=0.48\textwidth]{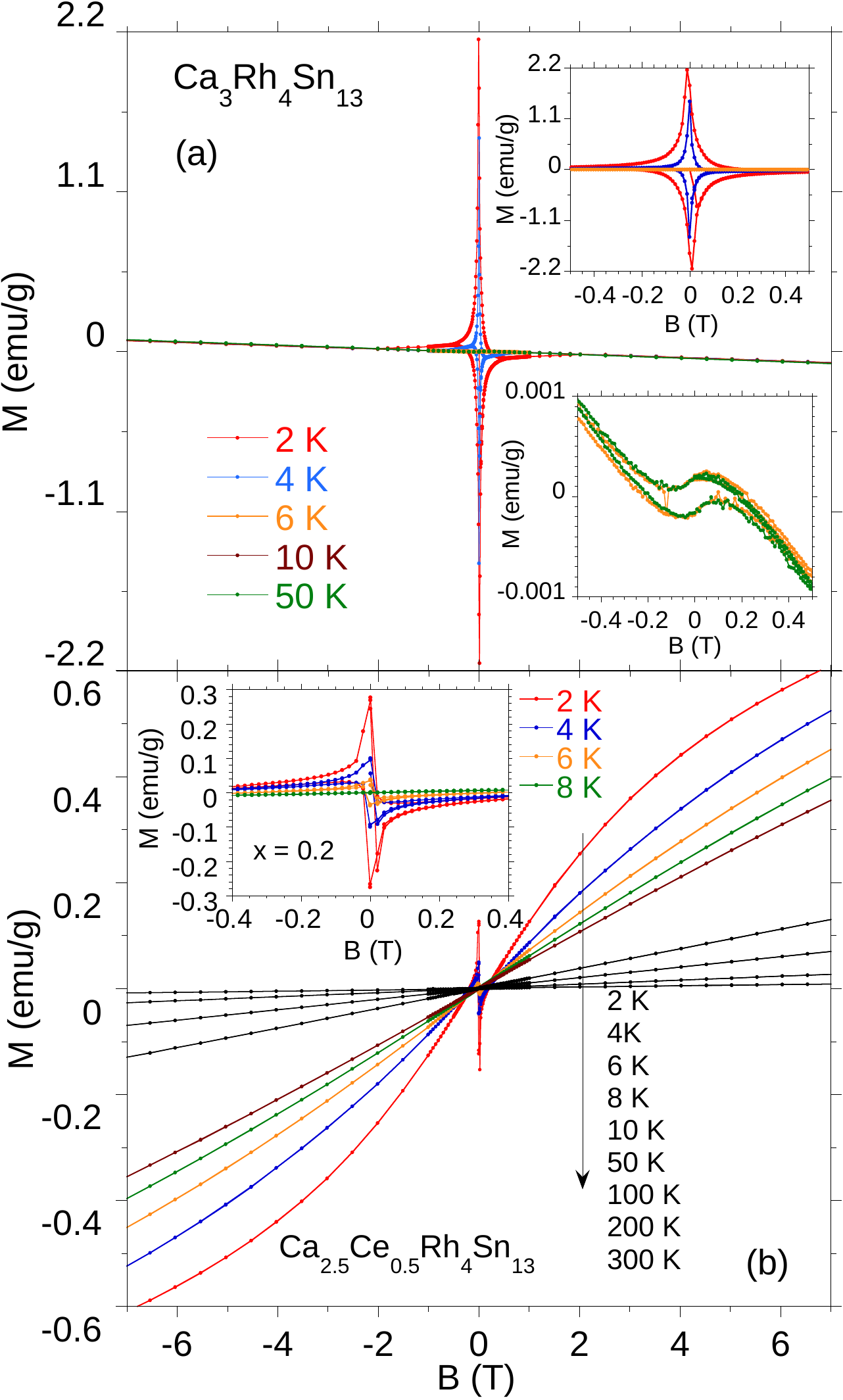}
\caption{\label{fig:Fig5}
Magnetization $M$ vs magnetic field $B$ for Ca$_{3}$Rh$_{4}$Sn$_{13}$ ($a$) and Ca$_{2.5}$Ce$_{0.5}$Rh$_{4}$Sn$_{13}$ ($b$) at different temperatures. The insets exhibit hysteresis loops at different temperatures.
}
\end{figure}

Figure \ref{fig:Fig5} shows the magnetization loops for the Ca$_{3-x}$Ce$_x$Rh$_{4}$Sn$_{13}$ samples with $x\leq 0.5$ and the irreversibility in the magnetization curves observed for $T<T_c$ at lower fields represents the effect of vortex pinning. Moreover, Ca$_{3}$Rh$_{4}$Sn$_{13}$ is diamagnetic in the wide temperature region (Fig. \ref{fig:Fig5}$a$)
with weak magnetization hysteresis loops observed at $T=6, 10$, and 50 K, which correlates with thermal hysteresis in the ZFC and FC $\chi$ data presented in Fig. \ref{fig:Fig2}$a$ and is not associated with the flux pinning effect. For the alloys doped with Ce  (cf. Fig. \ref{fig:Fig5}$b$), the magnetization $M$ vs $B$ isotherms for $B \gtrsim 1$ T are characteristic of paramagnets. The smaller hysteresis loop effect, as shown in the lower panel of Fig. \ref{fig:Fig5}, is likely due to atomic disorder and related to the more complicated effect of magnetic flux pinning.

\begin{figure}[h!]
\includegraphics[width=0.48\textwidth]{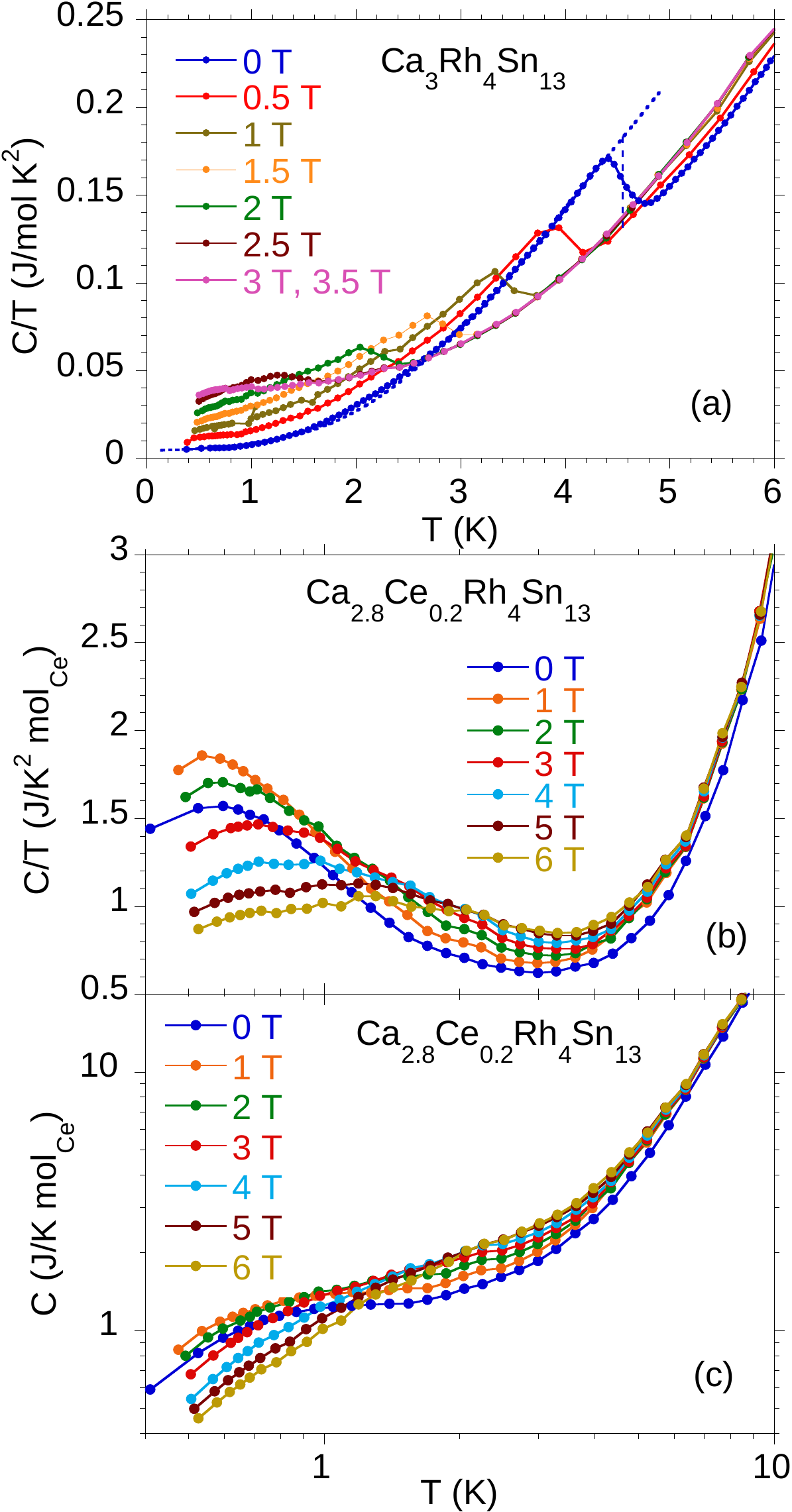}
\caption{\label{fig:Fig6}
($a$) Temperature dependence of specific heat, $C(T)/T$, for Ca$_{3}$Rh$_{4}$Sn$_{13}$ in various magnetic fields. The dotted blue line is the best fit of the expression $C(T)/T=\gamma +\beta T^2+\frac{1}{T}A\exp(-\Delta_c(0)/k_{\rm B}T)$ to the data. The $C(T)/T$ and $C(T)$ data vs $T$ at different magnetic fields are shown  for Ca$_{2.8}$Ce$_{0.2}$Rh$_{4}$Sn$_{13}$ in panel ($b$) and ($c$), respectively.
}
\end{figure}

A sharp superconducting transition in the specific heat data $C(T)/T$ at $T_c = 4.6$ K is shown for Ca$_{3}$Rh$_{4}$Sn$_{13}$ in Fig. \ref{fig:Fig6}$a$. The $C(T)/T$ data are well approximated by the expression $C(T)/T=\gamma +\beta T^2+\frac{1}{T}A\exp(-\Delta_c(0)/k_{\rm B}T)$. The best fit gives an electronic coefficient $\gamma=6.1$ mJ/mol K$^2$, $\beta=4$ mJ/mol K$^4$ and the energy gap at $T=0$, $\Delta_c (0)/k_{\rm B}=16.0$ K. From $\beta =N(12/5)\pi^4R\theta_D^{-3}$, we estimated the Debye temperature, $\theta_D \sim 213$ K, which compares very well with the literature data \cite{Haas1999}. 
The $C/T \sim \exp(-\Delta_c(0)/k_{\rm B}T)$ exponential behavior indicates an $s$-wave and BCS character of superconductivity. We obtained $\Delta C/(\gamma T_c)\cong 1.4(7)$ based on a value of the electronic specific heat coefficient $\gamma = 34$ mJ/mol K$^2$ for $T$ $\leq$ $T_c$ at a field of 3 T; we also determined $2\Delta_c (0)/k_{\rm B}T_c \approx 7.0$ which is larger than that expected from the BCS theory ($2\Delta_c (0)/k_{\rm B}T_c=3.52$) which indicates that Ca$_{3}$Rh$_{4}$Sn$_{13}$ may be categorized as a strong coupling superconductor.
The two lower panels, Fig. \ref{fig:Fig6} ($b$) and ($c$) display the specific heat data $C(T)/T$ and $C(T)$, respectively, for Ca$_{2.8}$Ce$_{0.2}$Rh$_{4}$Sn$_{13}$. 
There is no sharp transition at $T_c$ in the specific heat data shown in panel ($b$) 
Instead, the specific heat displays a broad peak at $T\approx 0.6$ K  with a maximum value of $C/T = 1.9$ J/K$^2$ mol$_{Ce}$ which is strongly reduced by field, shifts to higher temperatures, and is not related to the superconductivity of the sample. We attribute this low-temperature $C$ behavior to the formation of a spin-glass-like magnetic state. 
The low-$T$ heat capacity gives magnetic entropy $S=1$ J/K mol$_{Ce}$, while the limit $S=R\ln 2$ \cite{comment1} is reached at about 6 K, well above the $C/T$ maximum. 
If the cause of the maximum in $C$ and $C/T$ was due solely to a Kondo effect, then  $C/T$ should be about 4 J/K$^2$ mol$_{Ce}$ \cite{Slebarski2014b} in the limit of $T\longrightarrow 0$ 
(
recently we have shown a single ion Kondo impurity state for La-diluted isostructural Ce$_{3-x}$La$_{x}$Co$_{4}$Sn$_{13}$ \cite{Slebarski2014b} and Ce$_{3-x}$La$_{x}$Rh$_{4}$Sn$_{13}$ \cite{Slebarski2014}
series of compounds  with small value of Kondo temperature of about 1.5 K and $C(T)/T$ value almost not $x$-dependent and about $3-4$ J/K$^2$ mol$_{Ce}$.)
Moreover, the $4f$ contribution to the specific heat, $\Delta C(T)$, should be approximated by the Kondo resonant-level model \cite{Schotte75}, however, this is not a case. 
Within this model the Kondo-impurity contribution to $\Delta C$ can be described by the expression:
\begin{multline}
\Delta C=R \frac{2S\Delta_K}{\pi k_{\rm B}T}-2RRe\bigg\{\frac{(\Delta_K +ig\mu_{\rm B}H)^2}{(2\pi k_{\rm B}T)^2} \\  
\times \bigg[(2S+1)^2 \psi^{'}\bigg(1+\frac{\Delta_K +ig\mu_{\rm B}H}{2\pi k_{\rm B}T}(2S+1)\bigg) \\ 
- \psi^{'}\bigg(1+\frac{\Delta_K +ig\mu_{\rm B}H}{2\pi k_{\rm B}T}\bigg)\bigg]\bigg\},
\end{multline}
where $\psi^{'}$ is the first derivative of the digamma function, and $\Delta_K/k_{\rm B}$ is of the order of the Kondo temperature $T_K$.
The best approximation gives $\Delta_K=0$, which eliminates the Kondo impurity effect as dominant. Therefore, 
the $C(T)$ and $\chi_{ac}(T)$ data for Ca$_{2.8}$Ce$_{0.2}$Rh$_{4}$Sn$_{13}$ indicate the coexistence of an inhomogeneous superconducting phase
(cf. Ref. \onlinecite{Slebarski2014a,Slebarski2015}) and a spin-glass-like state at temperatures below $T_c$. 
Very recently we have discussed a very similar specific heat behavior in so strongly disordered La$_{3}$Co$_x$Ru$_{4-x}$Sn$_{13}$ superconductors \cite{Slebarski2015} that
the expected specific heat jump and the onset of diamagnetism are spread out over a very large temperature range. It has been shown that 
potential disorder smooth on a scale comparable to the coherence length leads to large modulation of the superconducting gap and large transition width, similar to that shown in Fig. \ref{fig:Fig4}. 

\begin{figure}[h!]
\includegraphics[width=0.48\textwidth]{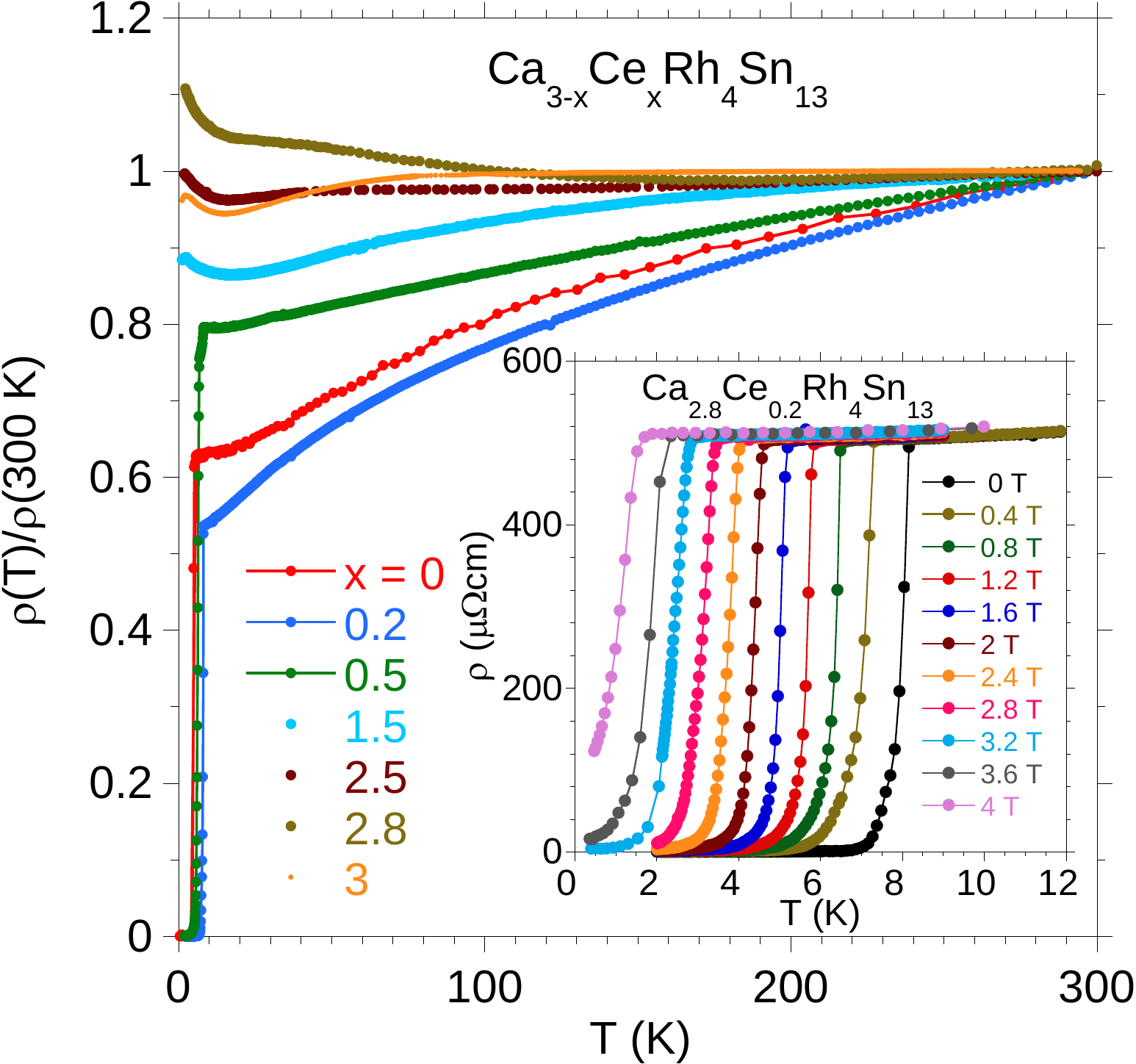}
\caption{\label{fig:Fig7}
Electrical resistivity $\rho (T)$ normalized by its room temperature value for Ca$_{3-x}$Ce$_x$Rh$_4$Sn$_{13}$ compounds. The inset shows the temperature dependence of $\rho$ for Ca$_{2.8}$Ce$_{0.2}$Rh$_{4}$Sn$_{13}$ at various externally applied magnetic fields.
}
\end{figure}

\begin{figure}[h!]
\includegraphics[width=0.47\textwidth]{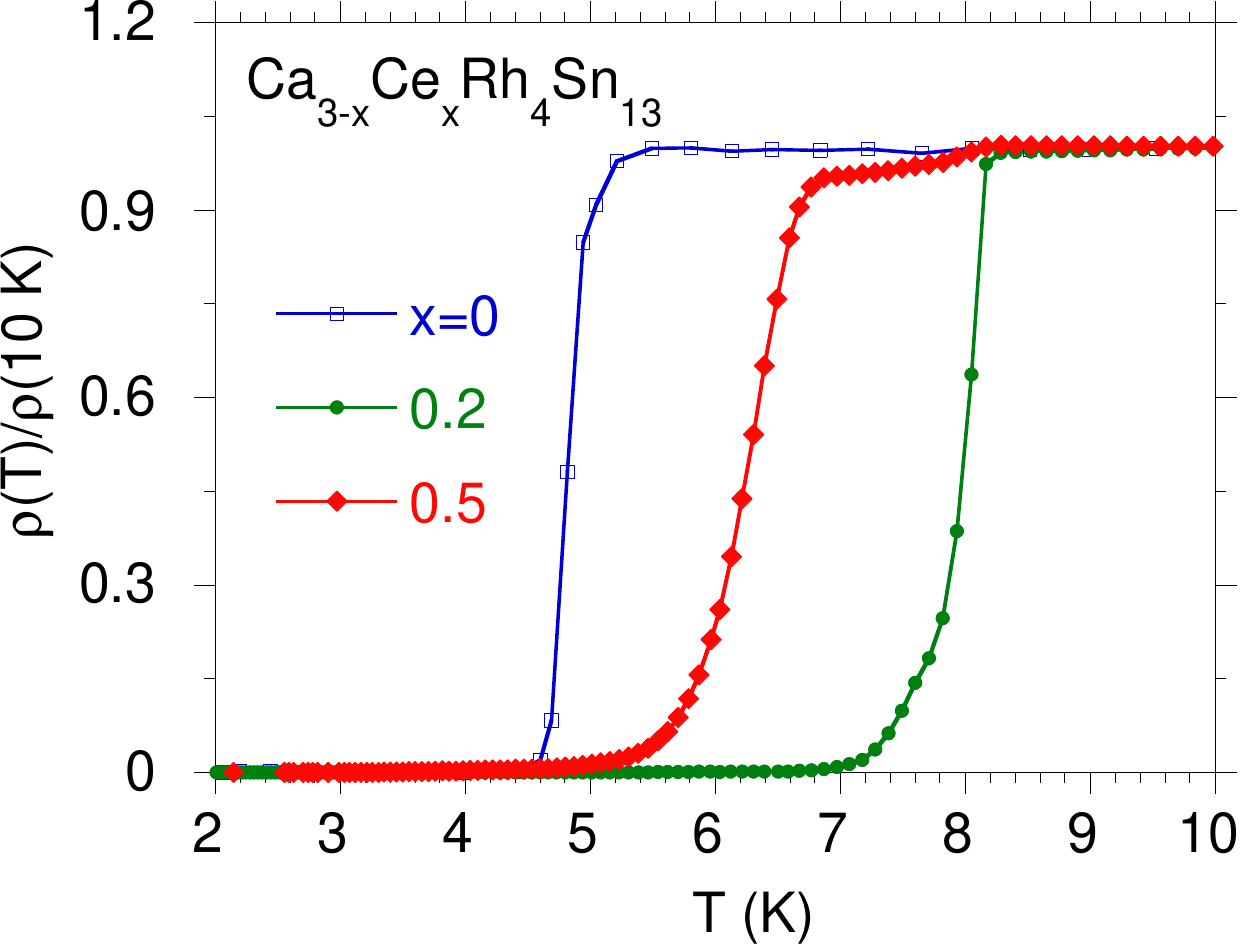}
\caption{\label{fig:rho_tc}
Electrical resistivity of the superconducting Ca$_{3-x}$Ce$_x$Rh$_4$Sn$_{13}$ samples ($x=0$, 0.2, and 0.5) normalized to $\rho$ at $T=10$ K, near the critical temperature $T_c$.}
\end{figure}

Figure \ref{fig:Fig7} displays the zero field electrical resistivity $\rho$ between 0.4 and 300 K for Ca$_{3-x}$Ce$_x$Rh$_4$Sn$_{13}$ compounds, normalized to the value of $\rho$ at $T=300$ K. The samples $0\leq x\leq 0.5$ show a superconducting transition. The remaining compounds in the series for which $x>0.5$ exhibit $\rho (T)$ that is characteristic of Kondo-lattice systems, with distinct anomalies at $T\approx 160$ K for the samples $x=2.5$ and 2.8 due to a subtle deformation of the Sn$_{12}$ cages (cf. Refs. \onlinecite{Slebarski2014b,Slebarski2013}, and Fig. \ref{fig:XRD}). Now we will discuss the $\rho(T)$ data for the superconducting samples. Figure \ref{fig:rho_tc} shows the temperature dependence of the elecrical resistivity for $x=0,\:0.2$ and 0.5. The transitions in the doped samples are much broader than for the $x=0$ compound, which shows that substitution of Ce for Ca introduces inhomogeneity into the system.
It is interesting to note the shape of $\rho(T)$ for $x=0.5$ which exhibits two clear separate drops in $\rho$, one near $T_{c1}=8$ K and another near $T_{c2}=6.7$~K.  
$T_{c1}$ coincides with the temperature of the single transition for $x \gtrsim 1$.
The two distinct drops in $\rho$ for $x=0.5$ indicate a double resistive phase transition to the superconducting state. Such a double transition is typical for inhomogeneous (granular) superconductors \cite{granular}, in which isolated superconducting {\it islands} are formed at the higher temperature, $T_{c1}$. Then, at the lower temperature, $T_{c2}$, a global phase coherence develops at which point $\rho$ $\rightarrow$ 0. The double transition can also be seen in the ac susceptibility (see Fig. \ref{fig:Fig4}b): $\chi'$ starts to diminish at $T = 8$ K, but significantly decreases only
for $T < 6$ K. Additionally, the anomaly in $\chi''$ seen for temperatures between 6 K and 8 K (the inset in Fig. \ref{fig:Fig4}b) can be interpreted as a signature of the double transition \cite{Nakatsuji2006}.

\begin{figure}[h!]
\includegraphics[width=0.48\textwidth]{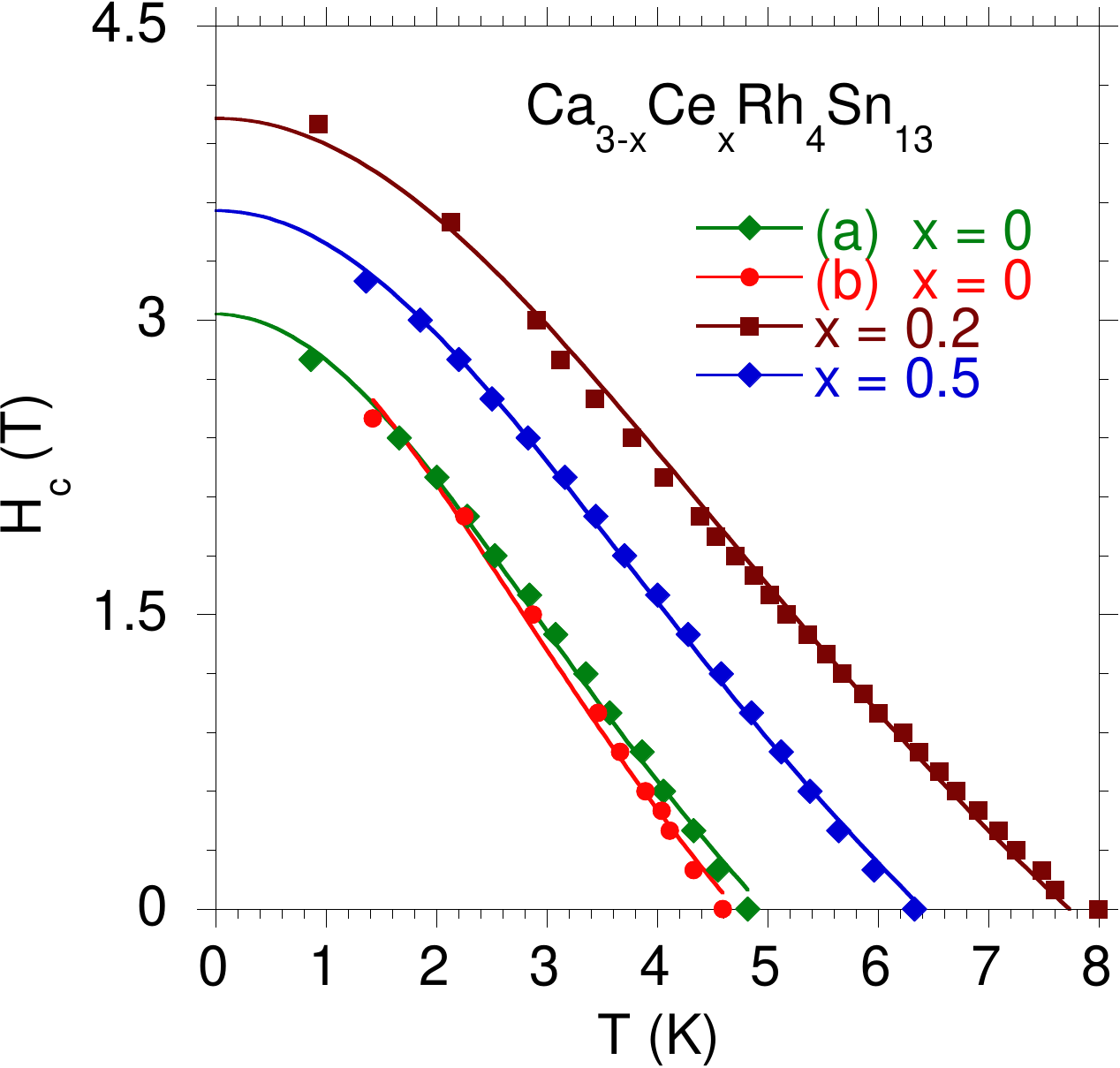}
\caption{\label{fig:Fig8}
Temperature dependence of the upper critical field $H_{c2}$ in the $H-T$ phase diagram for Ca$_{3}$Rh$_4$Sn$_{13}$ (symbols $a$ and $b$), Ca$_{2.8}$Ce$_{0.2}$Rh$_4$Sn$_{13}$, Ca$_{2.5}$Ce$_{0.5}$Rh$_4$Sn$_{13}$, respectively. $T_c$ values characterized by points $a$ for Ca$_{3}$Rh$_4$Sn$_{13}$, and for the samples $x=0.2$ and 0.5  are obtained from electrical resistivity data under $H$, and defined as the temperature at which $\rho$ drops to 50\% of its normal-state value. The red symbols $b$ represent $T_c$ obtained from $C(T)/T$ vs $T$ data in Fig. \ref{fig:Fig6}. The solid color lines represent a fit using the Ginzburg-Landau model of $H_{c2}(T)$.}
\end{figure}

The inset in Fig.~\ref{fig:Fig7} displays $\rho (T)$ for Ca$_{2.8}$Ce$_{0.2}$Rh$_4$Sn$_{13}$  in various magnetic fields. The $\rho (T)$ data exhibits a sharp transition near $T_c$ which is very similar to the behavior at $T_c$ that was observed for Ca$_{3}$Rh$_4$Sn$_{13}$ and Ca$_{2.5}$Ce$_{0.5}$Rh$_4$Sn$_{13}$. In Fig. \ref{fig:Fig8}, we show the $H-T$ phase diagram, where the critical temperatures were determined to be the temperature at which $\rho$ reaches 50\% of its value in the normal state. For the Ca$_{3}$Rh$_4$Sn$_{13}$ compound, there is also evidence of two superconducting phases, however, the nature of these transitions is different from that for Ca$_{2.5}$Ce$_{0.5}$Rh$_4$Sn$_{13}$. There is only one drop of resistivity, but its temperature is slightly off the position of the peak in the specific heat. A similar situation was observed in La$_{3}$Rh$_4$Sn$_{13}$ \cite{Slebarski2014a} and La$_{3}$Ru$_4$Sn$_{13}$ \cite{Slebarski2015}. The transitions are represented in Fig. \ref{fig:Fig8} by the $H-T$ curves ($a$) and ($b$); the data points along the ($a$) and ($b$) curves were obtained from the resistivity ($T_c^{\star}$ inhomogeneous phase) and specific heat ($T_c$ phase) data, respectively.
The $H-T$ data from all samples are well approximated by the Ginzburg-Landau (GL) theory.
The best fit of equation $H_{c2}(T)=H_{c2}(0)\frac{1-t^2}{1+t^2}$, where $t=T/T_c$ gives the upper critical field values of $H_{c2}(0)$ and $H^{\star}_{c2}(0)$  $\sim$ 3.1 T for Ca$_{3}$Rh$_4$Sn$_{13}$, $H^{\star}_{c2}(0)=3.6$ T for  Ca$_{2.8}$Ce$_{0.2}$Rh$_4$Sn$_{13}$, and $H^{\star}_{c2}(0)=4.0$ T for  Ca$_{2.5}$Ce$_{0.5}$Rh$_4$Sn$_{13}$.
The critical temperatures determined from the best fit of the GL equation to the $H-T$ plots are: $T_c=4.71$ K and $T_c^{\star}=4.97$ K for Ca$_{3}$Rh$_4$Sn$_{13}$, $T_c^{\star}=7.73$ K, and $T_c=6.41$ K for the samples substituted with nominal Ce concentrations of $x=0.2$ and $x=0.5$, respectively.
Using the Ginzburg-Landau relation \cite{Schmidt77} 
$\mu_0 H_{c2}(0)=\Phi_0/2\pi \xi(0)^2$ we determined the superconducting coherence length $\xi(0)$, where $\Phi_0=h/2e=2.068 \times 10^{-15}$ Tm$^2$ is the flux quantum. Ca$_{3}$Rh$_4$Sn$_{13}$ exhibits similar values of $\xi(0)$ and $\xi(0)^{\star}\cong 10.3 $ nm; for Ca$_{2.8}$Ce$_{0.2}$Rh$_4$Sn$_{13}$, $\xi(0)^{\star}\cong 9.6 $ nm, and for Ca$_{2.5}$Ce$_{0.5}$Rh$_4$Sn$_{13}$, $\xi(0)^{\star}\cong 9.0 $ nm. In Fig. \ref{fig:Fig8}, all these $H_c(T)$ curves have a small initial positive curvature. This feature can also be found in other skuterrudites, such as PrOs$_4$Sb$_{12}$\cite{Measson2004}. It is interesting that this compound also exhibits a double superconducting transition, but it was claimed in Ref. \onlinecite{Measson2008} that the positive curvature of $H_{c2}(T)$ is not related with sample inhomogeneities. A similar shape of $H_{c2}(T)$ also appears in MgB$_2$,\cite{MgB2} in borocarbides such as YNi$_2$B$_2$C and LuNi$_2$B$_2$,\cite{borocarbides} or in the cuprates.\cite{cuprates_Hc2}

\begin{figure}[h!]
\includegraphics[width=0.48\textwidth]{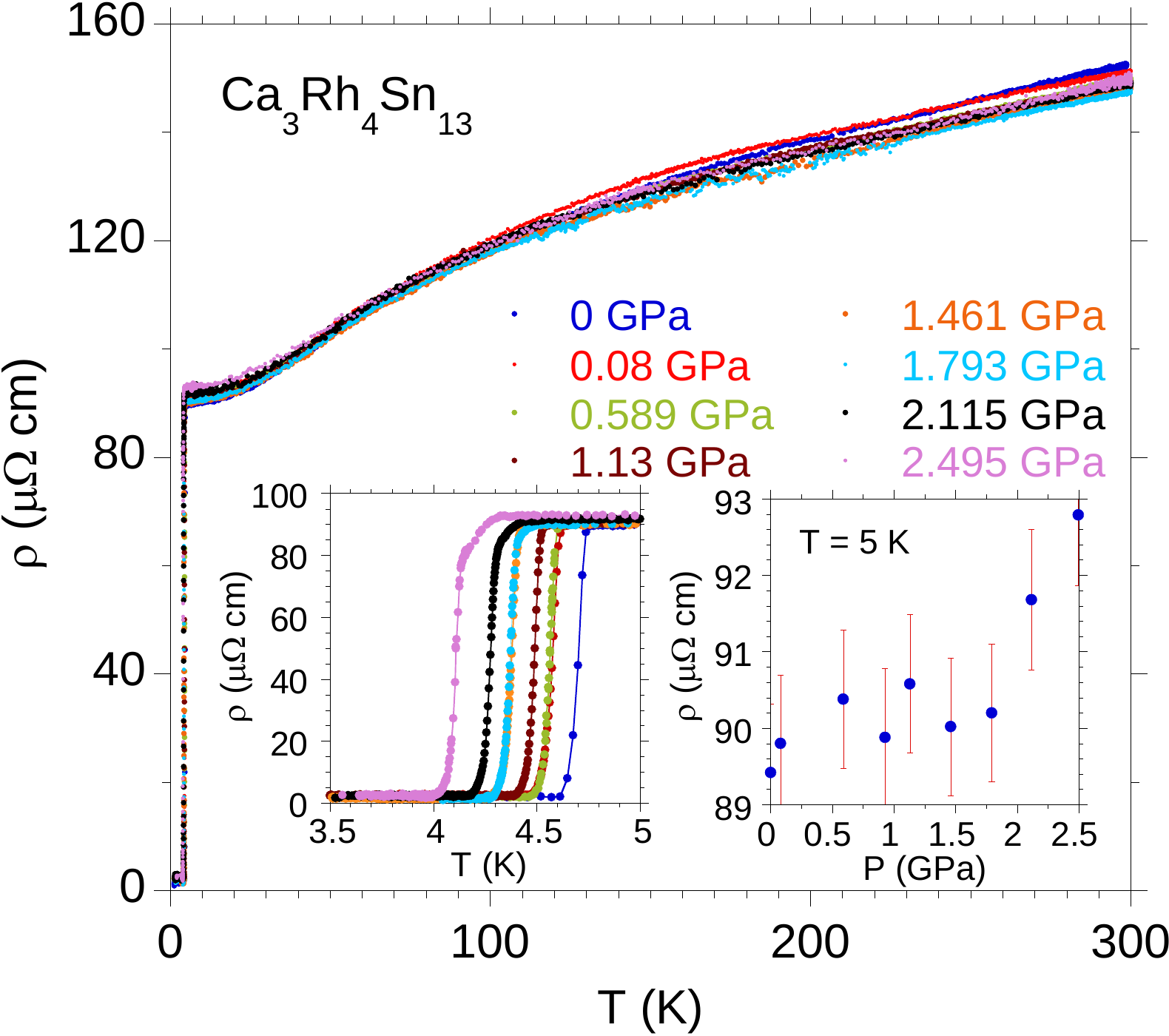}
\caption{\label{fig:Fig9}
Electrical resistivity for Ca$_{3}$Rh$_4$Sn$_{13}$ under applied pressure. The left inset shows the details near the critical temperature. The right inset displays the value of $\rho$ measured just above $T_c$ at $T=5$ K.
}
\end{figure}

\begin{figure}[h!]
\includegraphics[width=0.48\textwidth]{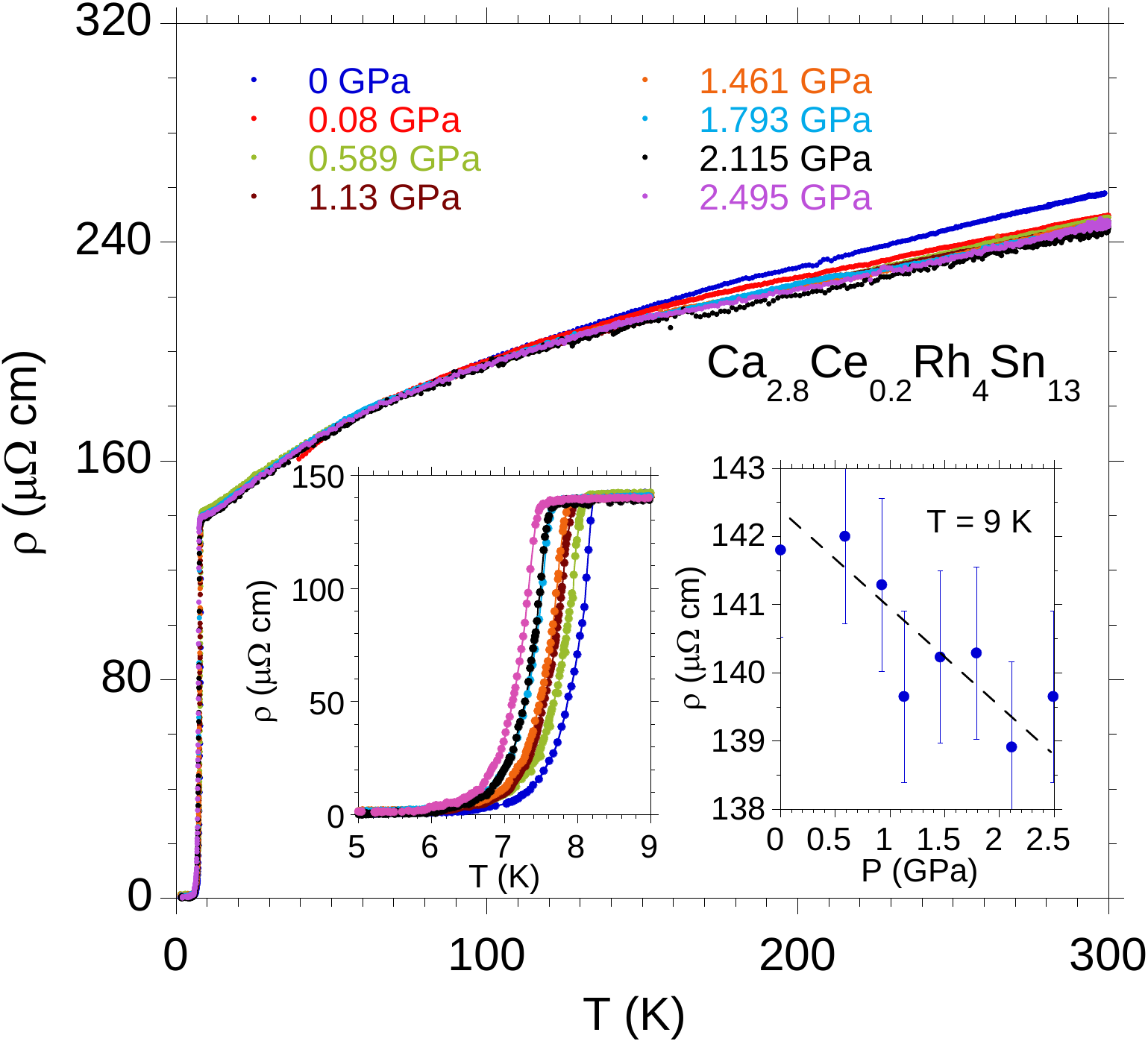}
\caption{\label{fig:Fig10}
Electrical resistivity for Ca$_{2.8}$Ce$_{0.2}$Rh$_4$Sn$_{13}$ under applied pressure. The left inset shows the details near the critical temperature. The right inset displays the value of $\rho$ measured just above $T_c$ at $T=9$ K.
}
\end{figure}

The pressure ($P$) evolution of the electrical resistivity as a function of temperature for Ca$_{3}$Rh$_4$Sn$_{13}$ and Ca$_{2.8}$Ce$_{0.2}$Rh$_4$Sn$_{13}$ is displayed in 
Figs. \ref{fig:Fig9} and \ref{fig:Fig10}, respectively. From these  data we obtained the pressure coefficients $\frac{dT_c}{dP}$ or $\frac{dT_c^{\star}}{dP}$ and $\frac{d\rho}{dP}$. ($i$) The pressure coefficients of $T_c^{\star}$ are $-0.2$ K/GPa for Ca$_{3}$Rh$_4$Sn$_{13}$ and $-0.3$ K/GPa for the $x=0.2$ sample substituted with Ce. These values of $\frac{dT_c^{\star}}{dP}$ are about twice those values for $T_c$, are found in similar isostructural La-based superconductors (cf. Ref. \onlinecite{Slebarski2015}), 
and  
seems to be  characteristic of materials which show evidence of nanoscale disorder leading to an inhomogeneous superconducting state with $T_c^{\star}>T_c$.

Within the Eliashberg theory of strong-coupling superconductivity, \cite{Eliashberg61}
the  McMillan expression, \cite{McMillan,Dynes72} 
\begin{equation}
T_c=\frac{\theta_\mathrm{D}}{1.45} \exp \left\{ \frac{-1.04(1+\lambda)}{\lambda - \mu^*(1+0.62\lambda)} \right\},              
\end{equation}
as a solution to the finite-temperature Eliashberg equations, gives an estimate of the electron-phonon coupling parameter $\lambda\approx 0.62$ (for $T_c^{\star}$ phase $\lambda^{\star}\approx 0.63$), when the Coulomb repulsion $\mu^{\star}$ is assumed to be $\sim 0.1$ as a typical value known for $s$ and $p$ band superconductors. 
The electron-phonon coupling $\lambda$ is given by the expression \cite{McMillan,Hopfield},
\begin{equation}
\lambda=\frac{N(E_{\rm F})\langle I^2 \rangle}{M\langle \omega^2 \rangle},
\end{equation}
where $\langle I^2 \rangle$ is the square of the electronic matrix element of electron-phonon interactions averaged 
over the Fermi surface, $\langle \omega^2 \rangle$ is an average of the square of the phonon frequency, and $M$ is  the atomic mass,
is larger for the inhomogeneous superconducting $T_c^{\star}$ state with respect to the bulk effect observed below $T_c$,  which may lead to a larger value of $\frac{dT_c^{\star}}{dP}$ than $\frac{dT_c}{dP}$. The primary reason for $\frac{dT_c^{\star}}{dP} > \frac{dT_c}{dP}$ is the pressure dependence of $\theta_D$, which leads to larger lattice stiffening in the $T_c^{\star}$ phase with respect to the bulk effect below $T_c$ and contributes to the $T_c^{\star}>T_c$ effect.
The dependence is given
by the Gr\"uneisen parameter $\gamma_G=-\frac{dln{\theta_D}}{dlnV}$, which determines the lattice stiffening. Using the McMillan expression it was found \cite{Shao2004} that $\gamma_G$ strongly determines the magnitude and sign of ${dT_c}/{dP}$. It is also probable that in the case of inhomogeneous superconductivity, the pressure dependence of
the density of states at the Fermi level is more pronounced than in bulk superconductors, and may lead to a larger value of
${dT_c^\star}/{dP}$ than ${dT_c}/{dP}$.
($ii$) For $T>T_c $, the electrical resistivity (Fig. \ref{fig:Fig9}) shows the positive coefficient $\frac{d\rho}{dP}$ for Ca$_{3}$Rh$_4$Sn$_{13}$, while for the Ce doped sample (Fig. \ref{fig:Fig10}) $\frac{d\rho}{dP}<0$. This diverse  $\rho$ vs $P$ behavior is interpreted as a result of the pressure dependent band structure near the Fermi energy (which will be discussed  below).

\subsubsection{Band structure of Ca$_{3}$Rh$_{4}$Sn$_{13}$ under pressure}

\begin{figure}[h!]
\includegraphics[width=0.48\textwidth]{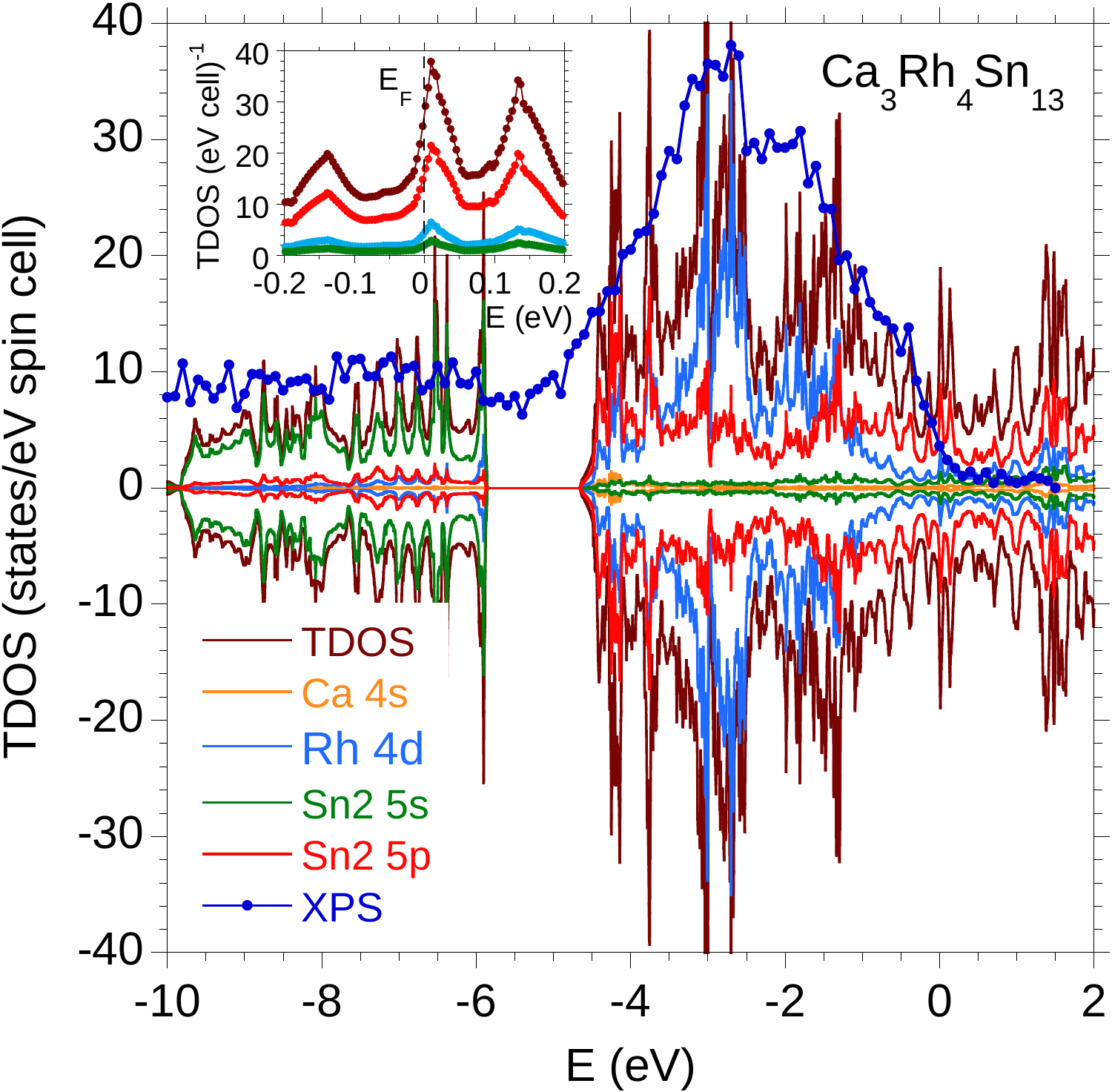}
\caption{\label{fig:Fig11}
Valence band XPS spectrum for Ca$_{3}$Rh$_4$Sn$_{13}$ compared with the calculated 
total and spin-resolved density of states  within the LSDA approximation. The figure also shows Ca $4s$, Rh $4d$, Sn2 $5p$ and $5s$ states for Ca$_{3}$Rh$_4$Sn$_{13}$. The inset exhibits the energy distribution of these partial DOS near the Fermi energy. 
} 
\end{figure}

Figure \ref{fig:Fig11}  compares the valence band X-ray photoelectron spectroscopy (VB XPS) spectra for Ca$_{3}$Rh$_4$Sn$_{13}$ and the calculated bands. The ground state is calculated as nonmagnetic. A comparison of the theoretical and experimental results shows that the calculated density of states (DOS) reflects all of the features found in the VB XPS spectra for Ca$_3$Rh$_4$Sn$_{13}$. The most intense peak at about 3 eV originates mainly from the Rh $4d$ states which are hybridized with the $5p$ states of Sn2. The inset exhibits a maximum in the DOS at the Fermi energy with a significant contribution coming from the Sn2 $5p$ states, which suggests the importance of these electronic states in electric transport. 

In order to understand the effect of pressure on the band structure, the DOS calculations were performed for hypothetical lattice parameters that are smaller compared to the parameters that were measured at room temperature. We used the Birch-Murnaghan  isothermal equation of state \cite{Murnaghan44} to estimate the hypothetical applied pressure that would correspond to a systematic decrease in the unit cell volume: 
$V(P)=V(0)[1+\frac{B^{'}}{B}P]^{-1/B^{'}}$, where $V(0)$ is the unit cell volume experimentally obtained at room temperature and ambient pressure, $B=91.64$ GPa  is the calculated bulk modulus, and $B^{'}=4.95$ is its pressure derivative. 
\begin{figure}[h!]
\includegraphics[width=0.48\textwidth]{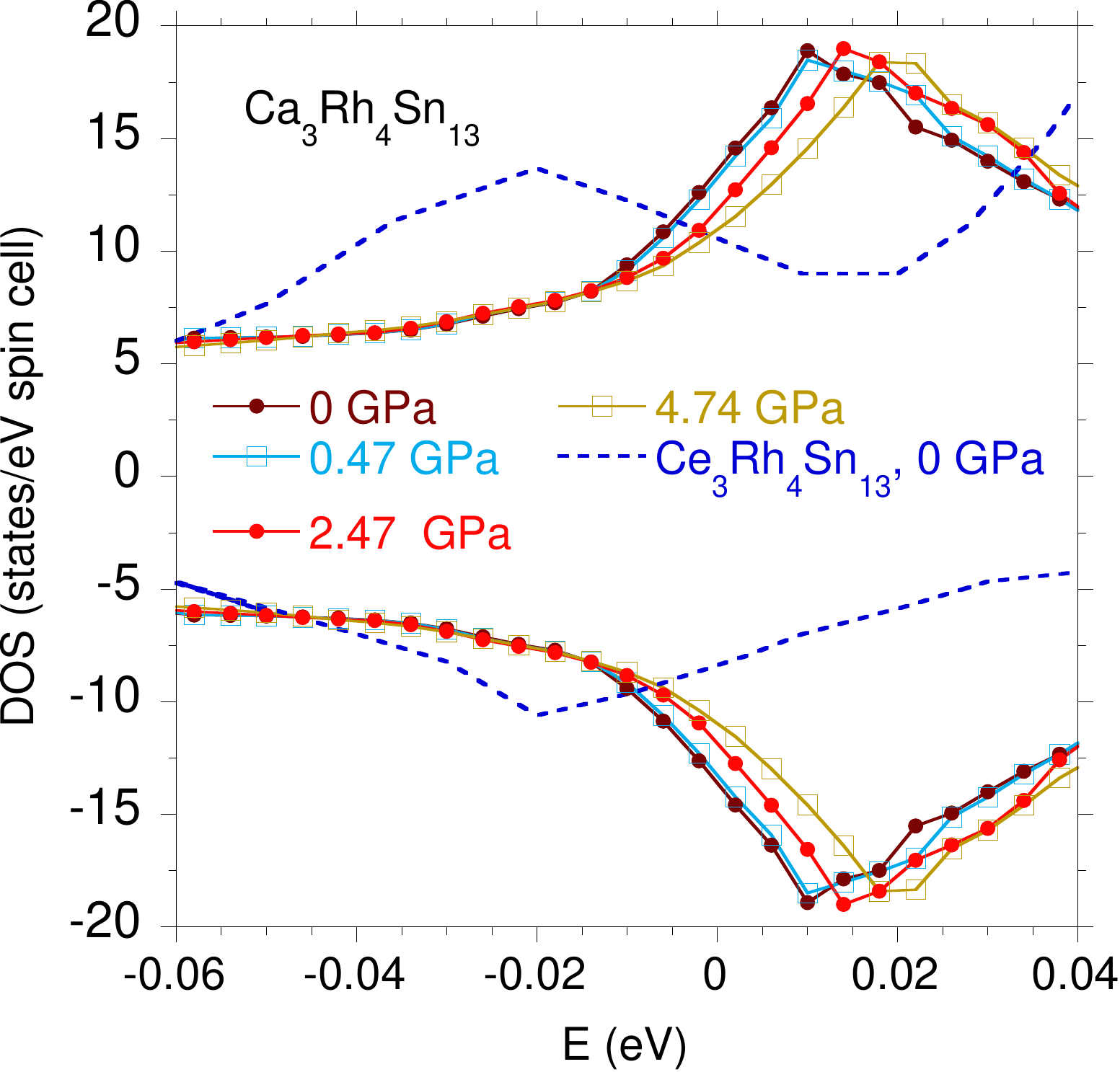}
\caption{\label{fig:Fig12}
The total and spin-resolved density of states near the Fermi energy within the LSDA approximation at varous pressures, calculated for Ca$_{3}$Rh$_4$Sn$_{13}$. The dotted blue line represents the total DOS calculated for Ce$_{3}$Rh$_4$Sn$_{13}$.  
}
\end{figure}
Figure \ref{fig:Fig12} displays the calculated DOS at various pressure. There is a systematic change in the DOS vs $P$ near $E_F$. However, it is interesting to note the change in the DOS vs $P$ at the Fermi energy as an explanation for the observed  coefficient $\frac{d\rho}{dP}>0$, shown for Ca$_{3}$Rh$_4$Sn$_{13}$ in Fig. \ref{fig:Fig9}. Our calculations documented the linear decrease of the total DOS at $E_F$ with $P$ in the region of $P<4.5$ GPa, giving $\frac{d(DOS)}{dP}\cong -1.3$ eV$^{-1}$ GPa$^{-1}$ 
(the proper DOSs taken from Fig. \ref{fig:Fig9}), 
which correlates well with the observed positive $\frac{d\rho}{dP}\cong 1.2$ $\mu\Omega$ GPa$^{-1}$ at applied pressure $P<2.5$ GPa. 
This simple estimate bases on the relation $\rho \sim 1/n$ between the resistivity and the number of carriers $n$, that naively reflects the DOS  at the Fermi level.
Moreover, the value of $\frac{d(DOS)}{dP}$  roughly agrees with the pressure dependence of $\rho$ in the limit of low $T$. 

Secondly, a linear change of the calculated DOS($P$) at $E_F$ with the change of the lattice parameter $a$ was obtained on the basis of the Birch-Murnaghan isothermal equation of state. Namely, the calculated DOS decreases with decreasing volume of the sample. Assuming that this linearity is extended to the region of {\it negative} lattice pressure (as is the case for Ca$_{3}$Rh$_4$Sn$_{13}$ with  $T_c\sim 8$ K, cf. Fig. \ref{fig:Fig1} and Ref. \onlinecite{Westerveld1987}) and that $U$ does not strongly depend on $P$, we attempt to demonstrate how the change in $DOS$ would increase  $T_c$ for Ca$_{3}$Rh$_4$Sn$_{13}$. This would follow from the BCS equation \cite{Bardeen57} $T_c=\theta_De^{-1/DOS(E_F)U}$ and the expression for the $DOS$: $DOS(E_F)U\sim \frac{\lambda - \mu^{\star}}{1+\lambda}$ \cite{Seiden1969}. We considered the calculated DOS for our sample   with $T_c=4.8$ K at ambient pressure in order to estimate the value of $U$.
A simple approximation gives $T_c\approx 10$ K, which is comparable to the value of $T_c\sim 8.4$ K that was found experimentally for the as-grown sample (cf. Fig. \ref{fig:Fig1} and Ref. \onlinecite{Westerveld1987}). 

The proposed model explains too the higher value of $T_c^{\star}=8$ K which was documented experimentally for the Ca$_{3-x}$Ce$_{x}$Rh$_4$Sn$_{13}$ samples with $0<x<1$ compared to the parent compound Ca$_{3}$Rh$_4$Sn$_{13}$ with $T_c^{\star}=4.9$ K. For Ca$_{2.8}$Ce$_{0.2}$Rh$_4$Sn$_{13}$, 
one can expect an increase of the DOS at $E_F$  caused by either the presence of Ce $f$-states (cf. Fig. \ref{fig:Fig12}  displays the total DOS for isostructural Ce$_{3}$Rh$_4$Sn$_{13}$) or a larger  lattice parameter (as is shown in Fig. \ref{fig:Fig1}) than what was observed in the parent compound. Within our model, even a slight increase in the DOS at $E_F$ may cause a significant increase in $T_c$ for
Ca$_{2.9}$Ce$_{0.1}$Rh$_4$Sn$_{13}$ relative to the value of $T_c$ observed for the parent compound Ca$_{3}$Rh$_4$Sn$_{13}$. It is apparent that the external pressure shifts the DOS of Ce$_{3}$Rh$_4$Sn$_{13}$ toward the Fermi energy and therefore one can expect an effective increase in the total DOS at $E_F$ with increasing $P$ for
Ce-doped alloys as well. This behavior correlates well with the observed negative $\frac{d\rho}{dP}$ at low $T$, as is shown in the inset of Fig. \ref{fig:Fig10}. 

Previously we discussed the effect of nanoscale disorder leading to an inhomogeneous superconducting state with the critical temperature $T_c^{\star}$ higher than $T_c$ of the bulk phase.  This scenario seems, however, not to be adequate for Ca$_{3}$Rh$_4$Sn$_{13}$, where the difference  $T_c^{\star}-T_c\sim 0.1$~K is small and hardness of the both superconducting phases is very similar. We can not, however, exclude the impact of the disorder on the $T_c^{\star}$ of Ce-doped alloy, as an additional effect, evident in temperature dependence of the specific heat.
It was shown theoretically \cite{Larkin1971,Galitski2002} and experimentally \cite{Young1999,Nadeem2015} that the superconducting transition temperature is higher in the presence of the spin-glass state compared to the noninteracting case. 
Generally, disorder can either suppress 
or significantly increase $T_c$ values 
\cite{Kuchinskii2015}.
The Ce-doped Ca$_{3}$Rh$_4$Sn$_{13}$ seems to be an example of strong-coupling superconductor with
the spin-glass-like state documented experimentally and generated by an atomic disorder.

\subsubsection{Bonding properties of Ce$_{3}$Rh$_4$Sn$_{13}$ investigated by electron localization function, absence of structural distortion}

For the series of skutterudite-related compounds, Ce$_3M_4$Sn$_{13}$ and La$_3M_4$Sn$_{13}$, where $M$ is a $d$-electron metal, the charge density analysis revealed a strong charge accumulation between the $M$ and Sn2 atoms, which implies a strong covalent bonding interaction and leads to a subtle structural transition at $T_D$ $\sim$ 150 K \cite{Slebarski2015a}.
In this class of materials, the structural deformation is usually accompanied by formation of a charge density wave (CDW) phase transition, and under external pressure $T_D\longrightarrow 0$ defines a novel structural quantum critical point \cite{Klintberg2012}. The structural deformation, however, was not documented for Ca$_{3}$Rh$_4$Sn$_{13}$. To determine the subtle bonding properties of 
the charge distribution in Ca$_3$Rh$_4$Sn$_{13}$ we present
a full-potential chemical-bonding analysis via calculation of the electron localization function  
within the density functional theory.\cite{Hohenberg1964} 
The ELF distribution is given by $\eta(\vec{r})=\left[1+\chi^2(\vec{r})\right]^{-1}$,  where
\begin{equation}
\begin{split}
\chi(\vec{r}) = \displaystyle\frac{\frac{1}{2}\sum_{i}\vert\vec{\nabla}\psi_i\vert^2 - \frac{1}{8} \frac{(\vec{\nabla}\rho)^2}{\rho}}{\frac{3}{10}(3\pi^2)^{2/3}\rho^{5/3}}.\
\end{split}
\end{equation}
$\vec{\nabla}\psi_i(\vec{r})\equiv \psi_{\nu \vec{k}}(\vec{r})$ are the crystal wave functions with band index $\nu$ and wavevector $\vec{k}$, and $\rho(\vec{r})$ is the total charge density. 
\begin{figure}[h!]
\includegraphics[width=0.48\textwidth]{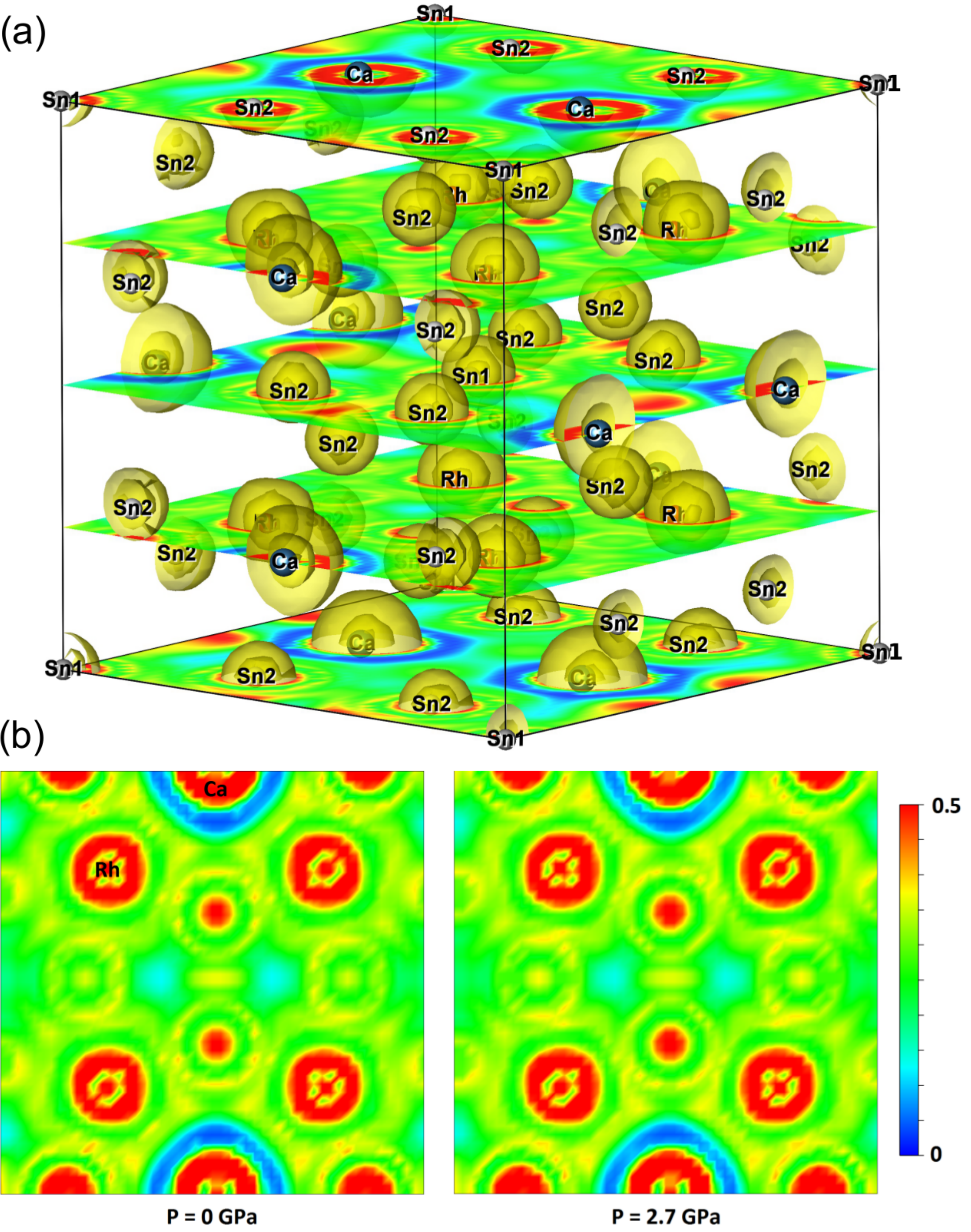}
\caption{\label{fig:Fig13}
($a$) ELF isosurfaces for Ca$_3$Rh$_4$Sn$_{13}$. ($b$) The ELF distribution in the plane (00$\frac{1}{4}$) for Ca$_3$Rh$_4$Sn$_{13}$ calculated for $P=0$ and 2.7 GPa. Note, that the pressure up to 2.7 GPa does not change the ELF distribution.
}
\end{figure}
ELF isosurfaces are shown in Fig. \ref{fig:Fig13}$a$. Figure \ref{fig:Fig13}$b$ exhibits the ELF distribution in the plane (00$\frac{1}{4}$) for Ca$_3$Rh$_4$Sn$_{13}$. The ELF maxima are located on the atoms in the plane, but the covalent bonding between Rh and Sn atoms, characteristic of  Ce$_3M_4$Sn$_{13}$ and La$_3M_4$Sn$_{13}$ counterparts \cite{Slebarski2015a} are not observed. This covalent bonding  is a cause of structural distortion at $T_D$ $\sim$ $100-160$ K in the Ce$_3M_4$Sn$_{13}$ and La$_3M_4$Sn$_{13}$ compounds. This superlattice transition that is connected to a CDW instability of the conduction electron system which is observed in most of these compounds is not observed
in Ca$_3$Rh$_4$Sn$_{13}$. For Ca$_3$Rh$_4$Sn$_{13}$, the ELF analysis also confirms that the thermal hysteresis present in $\chi$ as shown in Fig. \ref{fig:Fig2} is not of the CDW origin but results rather from the inhomogeneity of the sample.

\section{Electronic structure and magnetic properties of Ca$_{3-x}$Ce$_x$Rh$_{4}$Sn$_{13}$ with Kondo lattices for $x\geq 1.5$}

\begin{figure}[h!]
\includegraphics[width=0.48\textwidth]{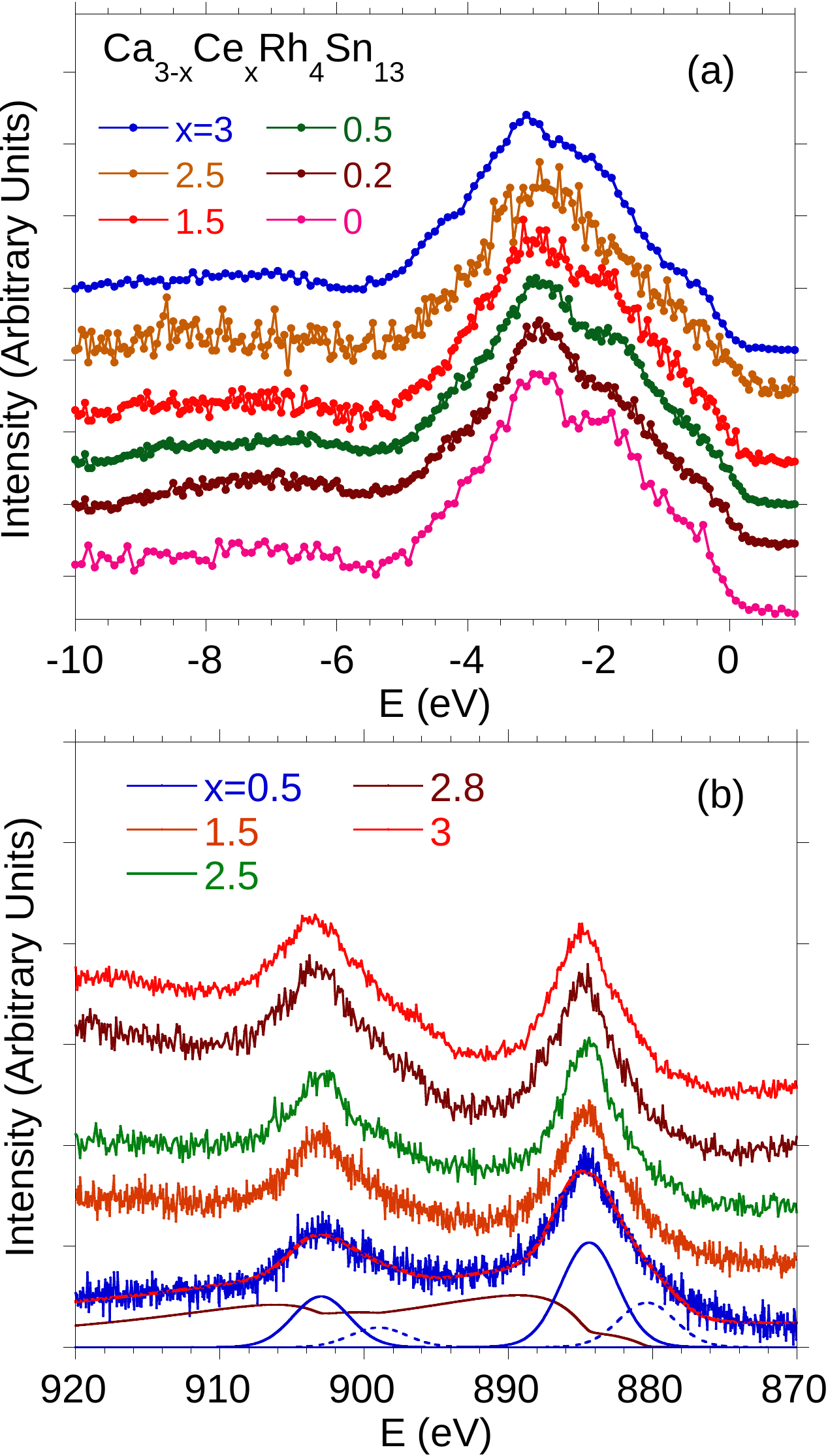}
\caption{\label{fig:Fig14}
($a$) Valence-band XPS spectra for Ca$_{3-x}$Ce$_x$Rh$_{4}$Sn$_{13}$, which show very similar structure across the series $x$. ($b$) The Ce $3d$ XPS spectra at the room temperature exhibit a deconvoluted spectrum for Ca$_{2.5}$Ce$_{0.5}$Rh$_{4}$Sn$_{13}$ on the basis of the Gunnarsson-Sch\"onhammer theoretical model,  as an example,  with spin-orbit components $3d^94f^0$ (blue solid line) and $3d^94f^2$ (blue dotted line). The brown solid line represents the background, the solid red line shows the fit after deconvolution to the XPS spectra. (From the deconvolution procedure, the Sn $3s$ line contribution at  885 eV  provides about 15 \%  of the total peak intensity due to $3d^9_{5/2}4f^1$ final states.)
} 
\end{figure}

In order to explain the (de)localized character of the Ce $4f$-electron states in the system of Ca$_{3-x}$Ce$_x$Rh$_{4}$Sn$_{13}$ alloys we analyze the valence band and the Ce-$3d$ core level XPS spectra. 
The VB XPS bands shown in Fig. \ref{fig:Fig14}$a$ are very similar across the series $x$ of compounds we investigated, which suggests that the Rh $4d$-electron states dominate the shape  of the bands, while the $4f$-electron state contribution is small.  This behavior is typical for isostructural compounds $RE_3M_4$Sn$_{13}$. 
The LSDA calculations (cf. Fig. \ref{fig:Fig11}) support these experimental observations.
In panel ($b$) of Fig. \ref{fig:Fig14}, the Ce-$3d$ core level XPS spectra demonstrate the well localized Ce $4f$-electron states and the weak hybridization between the Ce $4f$-electron and conduction band states, characterized by energy $\Delta$. 
The qualitative analysis of the $3d$ XPS spectra shown in Fig. \ref{fig:Fig14}$b$ was performed based on the Gunnarsson and Sch\"{o}nhammer theoretical method. The model  (for details see Refs. \onlinecite{Gunnarsson83,Fuggle83,Slebarski04}) explains 
a complex structure of the Ce-$3d$ core level XPS spectra in which the main $3d^9_{5/2}4f^1$ and $3d^9_{3/2}4f^1$ spin-orbit splitting components of the final states are associated with the stable configuration of the Ce-$4f$ shell and the contributions  
from the $3d^94f^0$ and $3d^94f^2$ are due to on-site hybridization between the $f$-electron states and conduction band. The presence of the $3d^94f^0$ sattelite line usually indicates an intermediate valence of the Ce ions, while $3d^94f^2$ reflects the hybridization effect which is expressed by energy $\Delta=\pi V^2 DOS(\epsilon_F)$. Standard analysis of the intensities of the components of the Ce-$3d$ XPS spectra shown in Fig. \ref{fig:Fig14}$b$ suggests both the presence of Ce$^{3+}$ ions 
\cite{comment3} and also a small hybridization energy $\Delta\approx 80\pm 20$ meV.

\begin{figure}[h!]
\includegraphics[width=0.48\textwidth]{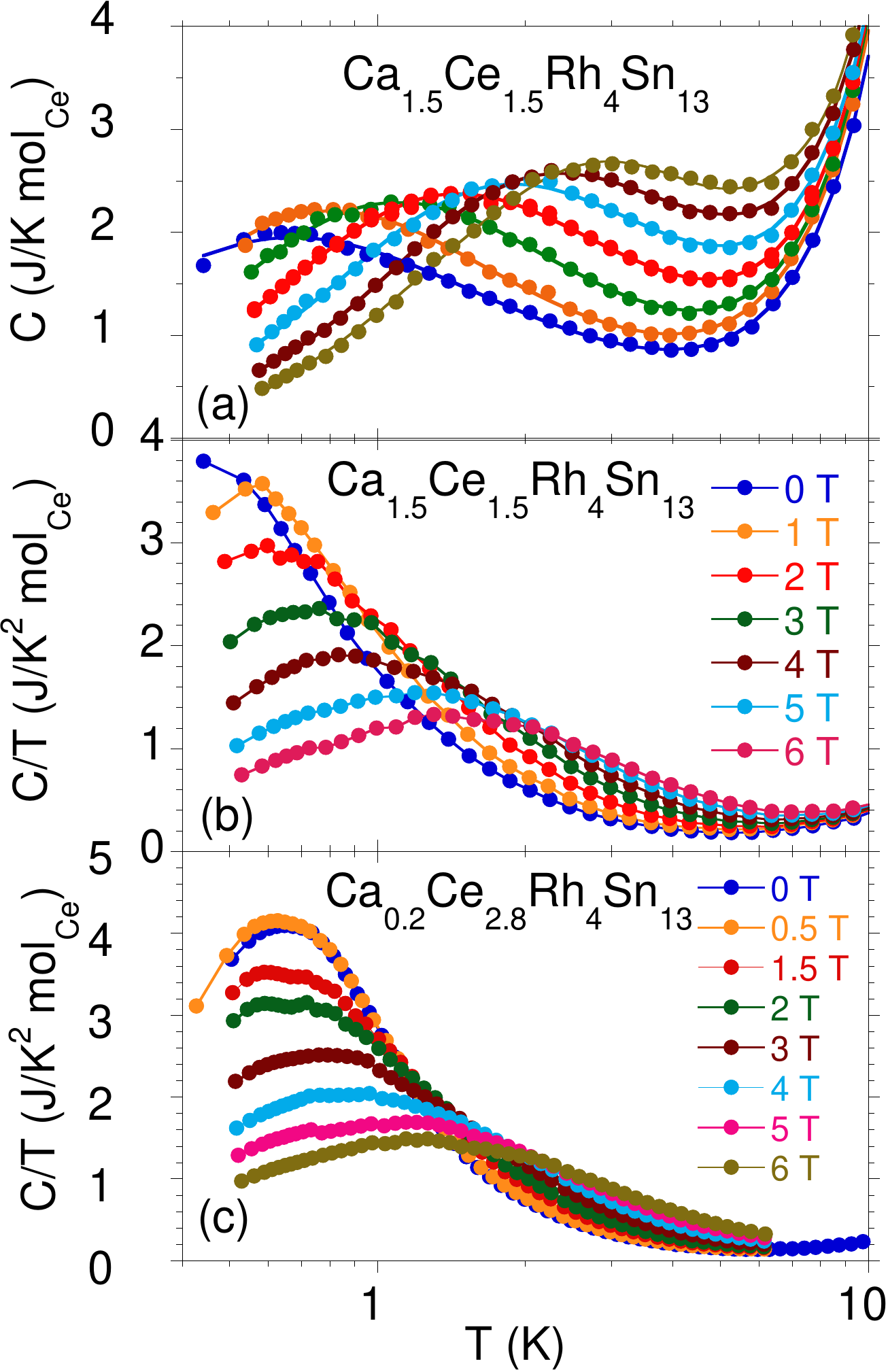}
\caption{\label{fig:Fig15}
($a$) $C(T)$ vs $T$ and ($b$) $C(T)/T$ vs $T$ for  Ca$_{1.5}$Ce$_{1.5}$Rh$_{4}$Sn$_{13}$. In panel ($a$) the lines represent the fit of the resonance level model to the experimental data. Panel ($b$) shows $C(T)/T$ at various magnetic fields for  Ca$_{0.2}$Ce$_{2.8}$Rh$_{4}$Sn$_{13}$.
} 
\end{figure}

\begin{figure}[h!]
\includegraphics[width=0.48\textwidth]{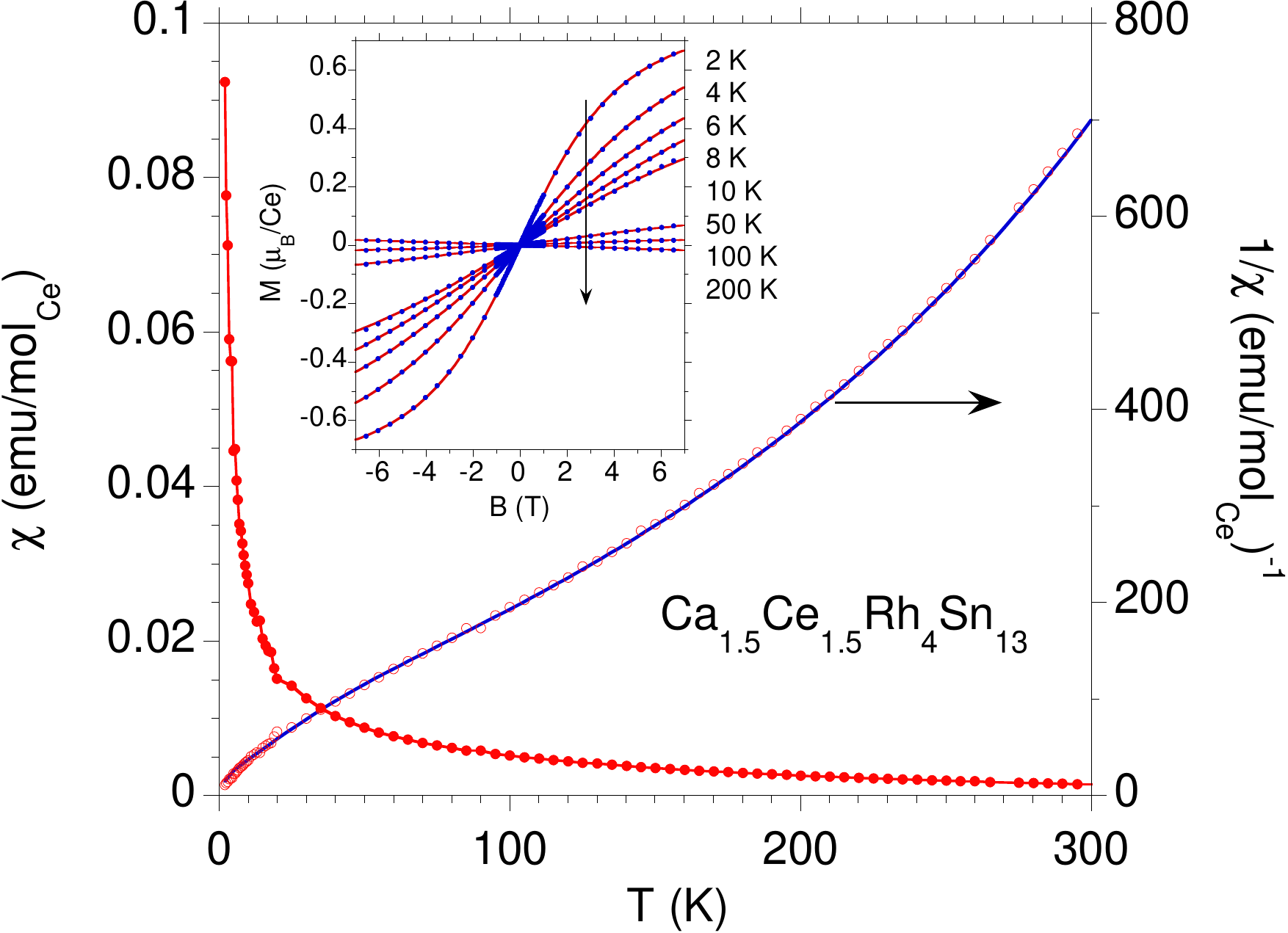}
\caption{\label{fig:Fig16}
Magnetic susceptibility $\chi$ and $1/\chi$ for Ca$_{1.5}$Ce$_{1.5}$Rh$_{4}$Sn$_{13}$ in an external field of 500 Oe. 
The blue line represents the CEF fit to the $1/\chi$ data, with
the two excited doublets separated from the ground state doublet by energy $\Delta_1=26$ K and $\Delta_2=303$ K, respectively.
Inset: Magnetization $M$ vs $B$ isotherms for Ca$_{1.5}$Ce$_{1.5}$Rh$_{4}$Sn$_{13}$ which are characteristic of paramagnets and well approximated by the Langevin function $L(\xi) = \coth(\xi)-1/\xi$, where $\xi =\mu B/k_{\rm B}T$ with a total magnetic moment $\mu \approx 0.9$ $\mu_{\rm B}$ at $T=1.8$ K. 
} 
\end{figure}

Our complex research suggests there is single-ion Kondo behavior in Ca$_{3-x}$Ce$_x$Rh$_{4}$Sn$_{13}$. Figure  \ref{fig:Fig15} shows the low-temperature  specific heat of Ca$_{1.5}$Ce$_{1.5}$Rh$_{4}$Sn$_{13}$ as $C(T)$ in panel ($a$)  and $C(T)/T$ in panel ($b$) as well as the low-temperature specific heat as $C(T)/T$ in panel ($c$) for the Ca$_{0.2}$Ce$_{2.8}$Rh$_{4}$Sn$_{13}$ compound. The large value of $C(T)/T \sim 4$ J/K$^2$ mol$_{Ce}$ for $T\rightarrow 0$ 
is typical for the family of  Ce$_3$M$_4$Sn$_{13}$ heavy Fermi liquids (c.f. Refs. \cite{Slebarski2014,Slebarski2014b}). Moreover, the $4f$-electron contribution to the specific heat $C$ is well approximated by the Kondo resonant-level model. In Fig. \ref{fig:Fig15}$a$ we present the fit of the expression $C(T)=\gamma_0 T+\beta T^3+ \Delta C$ to the experimental data, where $\Delta C$ is defined in Eq. 1. The level width $\Delta_K/k_{\rm B}\approx 0.95$ K for $B=0$ and is found to increase with magnetic field ($\Delta_K/k_{\rm B}\approx 2.25$ K for $B=6 T$). This behavior is characteristic of a single-ion Kondo system.
Figure \ref{fig:Fig16} shows the temperature dependence of the dc magnetic susceptibility $\chi $ and inverse susceptibility data, $1/\chi$, for Ca$_{1.5}$Ce$_{1.5}$Rh$_4$Sn$_{13}$, measured in a magnetic field of 500 Oe. 
The experimental $\chi (T)$ and $1/\chi (T)$ data can be well described in terms of CEF model \cite{Mulak86}:
\begin{equation}
\chi_{\rm CEF}=\frac{N\mu_{\rm B}^2}{k_{\rm B}}\frac{\sum_{i}(\frac{a_i}{T}+b_i)e^{-\beta\Delta_i}}{\sum_{i}g_i e^{-\beta\Delta_i}} + \chi_0.
\end{equation}
(Here, the summations run over all $i$ states of energies $E_i$ with $\Delta_i=E_i-E_0$ and Boltzman constant, $k_{\rm B}$). This model appropriately reflects the tetragonal Ce point symmetry, where the $J=5/2$ multiplet of the Ce$^{3+}$ ion splits into three doublets which are separated from the ground state by energies $\Delta_1\approx 26$ K and $\Delta_2\approx 303$ K, respectively, and $\chi_0\approx -0.0013$ emu/mol. 
The tetragonal symmetry of the CEF may indicate the structural deformation of the Sn2$_{12}$ cages in each of the Ca$_{1-x}$Ce$_{x}$Rh$_4$Sn$_{13}$ alloys. Recently \cite{Slebarski2013}, we documented that this subtle structural transition from the simple cubic structure at room temperature to a superlattice variant near $T =160$ K modifies the electronic structure and various physical properties of $RE_3$M$_4$Sn$_{13}$ compounds, where $RE=$ La or Ce and $M$ is a $d$-electron-type metal. Indeed, for the compounds in the Ca$_{1-x}$Ce$_{x}$Rh$_4$Sn$_{13}$ series with $x>1$, the measurements of electrical resistivity reveal an abnormal change at $T_D$ $\sim$ 160 K, which could be interpreted as a transition between the semi-metallic ($T<T_D$) and metallic ($T>T_D$) state, generated by a distortion of Sn2$_{12}$ cage. This interpretation is very probable, with consideration that the structural distortion was also found in the parent compound Ce$_{3}$Rh$_4$Sn$_{13}$.
In this case,
the Ca$_{1-x}$Ce$_{x}$Rh$_4$Sn$_{13}$ series appears to have a critical concentration $x_c\approx 1$, which separates the different behaviors of Ca$_{3}$Rh$_4$Sn$_{13}$ doped with Ce.
The inset to Fig.~\ref{fig:Fig16} shows
the magnetization $M$ vs $B$ isotherms for Ca$_{1.5}$Ce$_{1.5}$Rh$_4$Sn$_{13}$ which are characteristic of paramagnets.  They are well approximated by the Langevin function $L(\xi) = \coth(\xi)-1/\xi$, where $\xi =\mu B/k_{\rm B}T$, with the total magnetic moment $\mu \approx 0.9$ $\mu_{\rm B}$ at  $T=1.8$ K. The $M$ vs $B$ isotherms do not show any hysteresis in the field dependence of $M$ and we noted that the magnetization $M$ is well approximated by the Langevin function for
all the components of the Ca$_{1-x}$Ce$_{x}$Rh$_4$Sn$_{13}$ series with $x>1$.
\begin{figure}[h!]
\includegraphics[width=0.48\textwidth]{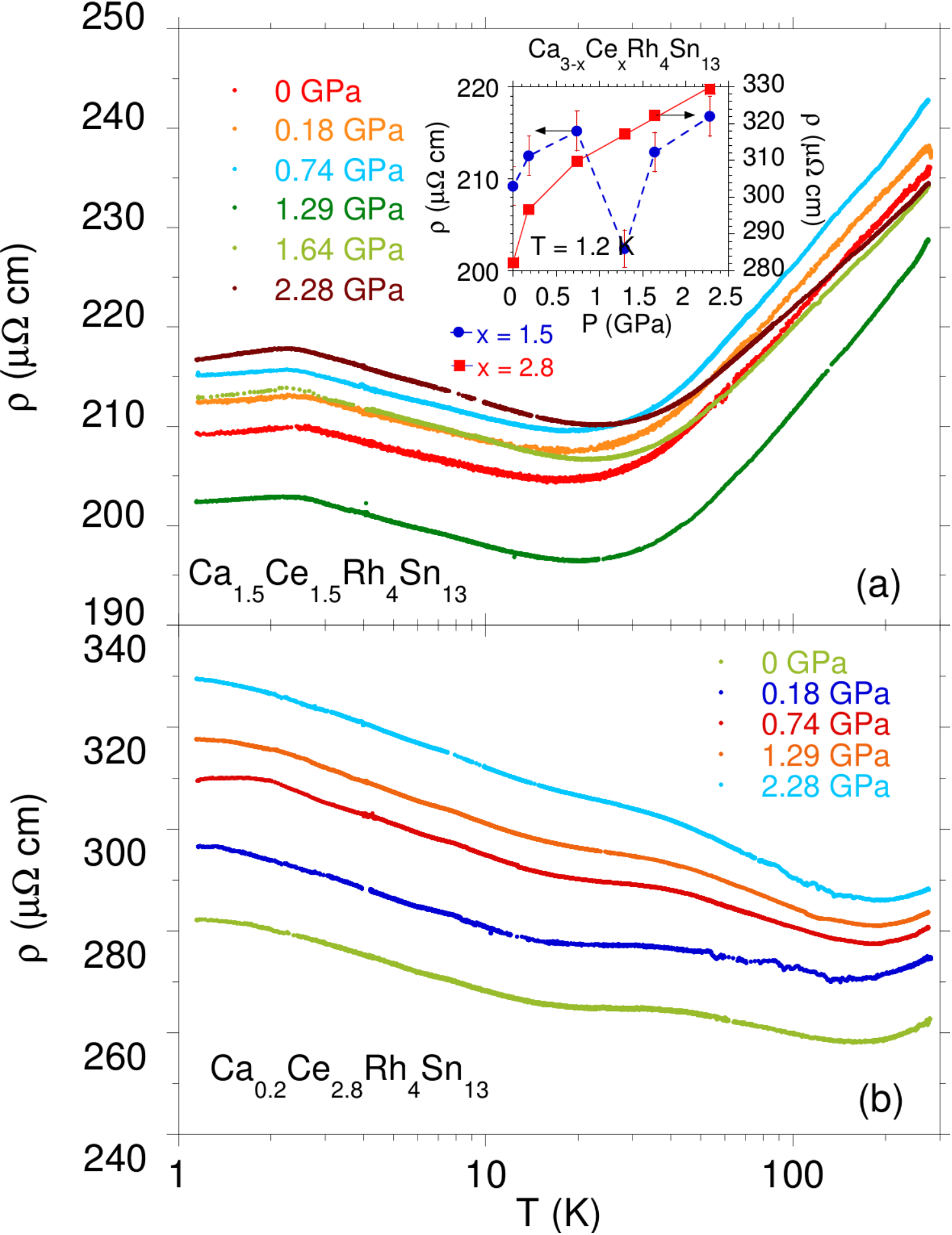}
\caption{\label{fig:Fig17}
Electrical resistivity for ($a$) Ca$_{1.5}$Ce$_{1.5}$Rh$_4$Sn$_{13}$
and ($b$) Ca$_{0.2}$Ce$_{2.8}$Rh$_4$Sn$_{13}$ under applied pressure. Inset: Electrical resistivity at $T= 1.2$ K as a function of $P$.
} 
\end{figure}
\indent Finally, in Fig. \ref{fig:Fig17} we present the electrical resistivity of Ca$_{1.5}$Ce$_{1.5}$Rh$_4$Sn$_{13}$ ($a$) and Ca$_{0.2}$Ce$_{2.8}$Rh$_4$Sn$_{13}$ ($b$) under applied external pressure. The pressure effect $\frac{d\rho}{dP}$ is strong and positive 
(in the upper panel the resistivity at $P=1.29$ GPa for the sample $x=1.5$ deviates from the expected value, this effect is not discussed here. The subtle DOS effect at $\epsilon_F$ could be a possible reason of this abnormal behavior, it seems to be also possible that this is not a physical effect).
A similar $\frac{d\rho}{dP}>0$ behavior was recently observed for Ce$_3$Co$_4$Sn$_{13}$ and Ce$_3$Rh$_4$Sn$_{13}$ \cite{Slebarski2015a} below the temperature of structural distortion and has been documented as a result of the band-structure properties near the Fermi energy under applied pressure.  

\section{conclusions}

The resistivity of Ce-substituted Ca$_3$Rh$_4$Sn$_{13}$ samples exhibits {\it high-temperature} superconductivity with the highest  $T_c\approx 8$ K for Ca$_{2.8}$Ce$_{0.2}$Rh$_4$Sn$_{13}$. This is much higher than the value of $T_c=4.8$ K for the parent compound. Magnetic measurements as well as theoretical band structure calculations predict a nonmagnetic ground state in Ca$_3$Rh$_4$Sn$_{13}$. It is possible, however, that the substitution of Ce for Ca forms a spin-glass-like phase which coexists with the superconducting phase and that the superconductivity is enhanced in this magnetic state. 
Although attempts have been made to explain the increase in $T_c$ for unconventional superconductors exhibiting disorder or a magnetic phase, there is no clear answer as to why $T_c$ rises in materials in which a spin-glass-like state coexists with superconductivity. The basis for the interpretation of the results presented here is that the Ce substitution into the superconducting parent compound Ca$_3$Rh$_4$Sn$_{13}$ 
increases the inhomogeneity and chemical pressure in the sample. It is also possible that the substitution affects $T_c$ through the increase of the DOS at the Fermi level. But results of the LSDA calculations presented in Fig. \ref{fig:Fig12} do not support such a scenario: the DOS at the Fermi level for Ce$_3$Rh$_4$Sn$_{13}$ is slightly smaller than for Ca$_3$Rh$_4$Sn$_{13}$, though it does not exclude the possibility that the DOS is larger at a lower doping level. It is also difficult to definitely say how the substitution modifies other parameters that enter the McMillan expression for $T_c$.
The Ce-doping drives Ca$_3$Rh$_4$Sn$_{13}$ through a $T_c$ vs $x$ superconducting  {\it dome} between $x=0$ and $x\approx 1.2$, similar to that documented for high-$T_c$ cuprates, and for $x>1.2$, the compounds settle into a Kondo-lattice state with structural distortion. 
Taken together, the resistivity, specific heat and susceptibility measurements are suggestive of granular superconductivity, a form of inhomogeneous superconductivity. Interestingly, a {\it high} value of $T_c$ was observed upon Ce substitution but in the absence of the structural deformation, while in the Kondo-lattice range, the slight structural deformation is visible in the resistivity data. Within the superconducting {\it dome} region we documented, the relationship between the electronic structure and the resistivity under external pressure for $T >T_c$. Namely, we calculated a linear decrease of the DOS at $E_F$ with $P$ which correlates well with the measured negative coefficient $\frac{d\rho}{dP}$.
These observations raise questions regarding the sensitivity of Cooper pairing in Ca$_{3-x}$Ce$_x$Rh$_4$Sn$_{13}$ and similar skutterudite-related systems to the electronic structure, atomic disorder and bonding, which suggests further detailed investigation.

\section{acknowledgments}

The research was supported by National Science Centre (NCN) on the basis of Decision No. DEC-2012/07/B/ST3/03027. M.M.M. acknowledges support by NCN under grant DEC-2013/11/B/ST3/00824. P.W. acknowledges support by NCN under grant DEC-2015/17/N/ST3/02361.
High-pressure research at the University of California, San Diego, was supported by the National Nuclear Security Administration under the Stewardship Science Academic Alliance program through the U. S. Department of Energy under Grant Number DE-NA0001841. One of us (A.\'S.) is grateful for the hospitality at the University of California, San Diego (UCSD). 

\newpage


\begin{thebibliography}{99}
\bibitem{comment1}
We did not find structural distortion for Ca$_{3}$Rh$_{4}$Sn$_{13}$. 
Considering a similar situation for Ca$_{2.8}$Ce$_{0.2}$Rh$_{4}$Sn$_{13}$, it can be assumed that the crystal field ground state of Ce ion in the cubic symmetry is a doublet.
\bibitem{comment3}
The occupation number of the $4f$ shell, $n_f=1-\nu$, and $\nu\cong \frac{I(f^0)}{I(f^0)+I(f^1)+I(f^2)}$. In case of the Ca$_{3-x}$Ce$_x$Rh$_{4}$Sn$_{13}$  series, the $3d^94f^0$ line is not observed in the Ce $3d$ XPS spactra, this suggests a stable configuration Ce$^{3+}$ of the $4f$ shell.
\end{thebibliography}
\end{document}